\documentclass[floatfix,aps,prb,amsmath,twocolumn,amssymb,titlepage,superscriptaddress]{revtex4-1}
\usepackage{amsthm,amsmath,amsfonts,dsfont} 
\usepackage{amsmath}    
\usepackage{graphicx}   
\usepackage{verbatim}   
\usepackage{subfigure}  
\usepackage{hyperref}   
\usepackage[ngerman,english]{babel}
\usepackage[usenames, dvipsnames]{xcolor}
\usepackage{tikz}
\usetikzlibrary{plotmarks,arrows.meta,calc,decorations.markings,math,arrows.meta,decorations.pathreplacing,decorations.pathmorphing}
\usepackage[percent]{overpic}
\def\be{\begin{equation}}
\def\ee{\end{equation}}
\usepackage{braket}
\begin{document}

\begin{flushright}
YITP-18-36
\end{flushright}

\title{Robustness of Topological Order in the Toric Code with Open Boundaries}

\author{Amit Jamadagni}
\email{amit.jamadagni@itp.uni-hannover.de}
\affiliation{Institut f\"ur Theoretische Physik, Leibniz Universit\"at Hannover, Appelstra{\ss}e 2, 30167 Hannover, Germany.}
\author{Hendrik Weimer}
\affiliation{Institut f\"ur Theoretische Physik, Leibniz Universit\"at Hannover, Appelstra{\ss}e 2, 30167 Hannover, Germany.}
\author{Arpan Bhattacharyya}
\email{bhattacharyya.arpan@yahoo.com}
\affiliation{Center for Gravitational Physics, Yukawa Institute for Theoretical Physics, Kyoto University, Kyoto 606-8502, Japan.}
\affiliation{Department of Physics and Center for Field Theory and Particle Physics, Fudan University, 220 Handan Road, 200433 Shanghai, P. R. China.}

\begin{abstract}
We analyze the robustness of topological order in the toric code in an open boundary setting in the presence of perturbations. The boundary conditions are introduced on a cylinder, 
and are classified into condensing and non-condensing classes depending on the behavior of the excitations at the boundary under perturbation. For the non-condensing class, we see 
that the topological order is more robust when compared to the case of periodic boundary conditions while in the condensing case topological order is lost as soon as the perturbation
is turned on. In most cases, the robustness can be understood by the quantum phase diagram of a equivalent Ising model.
\end{abstract}
\maketitle

\section{Introduction}
One of the main challenges in constructing a quantum computer has been to protect its qubits from decoherence. To overcome this challenge focus has shifted to topological
quantum computation \cite{Freedman1998, Freedman2003, Nayak2008}, where the information is naturally protected against local perturbation. In topological quantum computation, 
the information is encoded in the topological degrees of the freedom of the underlying system, the main ingredients of which are ``anyons" which are certain type of topological 
excitations characterized by fractional or non-abelian statistics. These excitations can be found in certain topological phases of matter which are mainly characterized by long-range
or short-range entanglement and have ground state degeneracy, gapless edge states as the signatures 
\cite{Wen1989, Niu1990, Wen1990, Nayak2008}.

In this article, we investigate the robustness of the toric code Hamiltonian in the presence of open boundaries. 
Our interest in open boundary conditions is two-fold: On the one hand, possible experimental realizations of the toric code \cite{Duan2003,Micheli2006,Jackeli2009,Lu2009,Weimer2010,Nielsen2010,Weimer2011,Barreiro2011,Fowler2012,Weimer2013a,Becker2013,Sameti2017}
will be much easier to implement for open boundaries than on a torus. On the other hand, it has been known that boundaries play an important role in the context of classifying different phases of matter.
E.g., various insulators with different symmetry breaking property separated by gapped domain wall exhibit rich physics in the low energy regime compared to  a single insulating phase. 
With the introduction of domain walls in the context of topological order much has been studied on how two different topological phases can be connected with each other through a boundary \cite{Beigi2011, Kitaev2012}, 
not only allowing the classification of phases via anyon condensation\cite{Hung2015} but also realizing the domain wall as a means for quantum computation \cite{Cong2017, Wang2017, Yoshida2016}. More importantly, from 
a practical point of view, these open boundary topologies play an important role in realizing ideas experimentally, as we have to deal with a finite system and also more trivial topologies. 
A systematic classification of boundaries in the context of Quantum Double models, including the domain wall classification has been discussed in Ref.~\onlinecite{Beigi2011}.
The notion of boundaries have also been discussed in the context of Levin-Wen string net models\cite{Wen2005} in Ref.~\onlinecite{Kitaev2012} using the language of category theory. 

Therefore, from a practical point of view it is important to understand the robustness of topological order in an open boundary setting under external perturbation.
The robustness of toric code, in a periodic boundary setting, has been extensively studied under various perturbations \cite{Nayak2007, Hamma2008, Dusuel2011, Chamon2008}. In this paper we address the 
robustness in the context of open boundaries given its importance as discussed above. We begin with a brief introduction of the toric code in a periodic boundary setting in Sec.~\ref{s2}, 
we review the open boundary conditions in Sec.~\ref{s3} and consider them in the presence of perturbation. We also review the map between the toric code under perturbation in an 
open boundary setting to an equivalent Ising model. In Sec.~\ref{s4}, using the above equivalence we gain an insight into the robustness of topological order.
In Sec.~\ref{s5}, we extend the discussion of Sec.~\ref{s4}, by numerically studying various signatures of topological order in the exact model. We observe that
Topological Entanglement Entropy acts as a good signature in detecting the phase transition, while a method based on Minimally Entangled States fails to 
capture the transition.

\section{Topological Order in the Toric Code \label{s2}}
\begin{figure}[t!]
\begin{tikzpicture}[scale=0.6]
\draw[line width=2pt, fill=LimeGreen](3,6)--(4,7)--(5,6)--(4,5)--(3,6)--cycle;
\draw[line width=2pt, fill=ProcessBlue](7,2)--(6,3)--(7,4)--(8,3)--(7,2)--cycle;
\draw[line width=2pt, LimeGreen](2,8)--(4,8);
\draw[line width=2pt, ProcessBlue](7,5)--(7,7);
\draw[step=2cm,black,very thin] (0.1,0.1) grid (9.9,9.9);
\foreach \x in {1,3,5,7} 
    \foreach \y in {2,4,6,8}
	\fill[Brown] (\x,\y) circle (5pt);
\foreach \x in {2,4,6,8} 
    \foreach \y in {1,3,5,7}
	\fill[Brown] (\x,\y) circle (5pt);
\draw[mark=square*,mark size=5pt,mark options={color=Red}] plot coordinates {(2,8)};
\draw[mark=square*,mark size=5pt,mark options={color=Red}] plot coordinates {(4,8)};
\node[scale=1.5] at (3.,8.5) {$i$};

\node[scale=1.5] at (7.5,5.5) {$j$};
\draw[mark=square*,mark size=5pt,mark options={color=Orange}] plot coordinates {(7,5)};
\draw[mark=square*,mark size=5pt,mark options={color=Orange}] plot coordinates {(7,7)};

\node[scale=1.5] at (4.25,5.75) {$v$};
\node[scale=1.5] at (7,3) {$p$};
\end{tikzpicture}
\caption{\label{fig1} Toric code model. The $A_{v}$ operator, denoted by the green diamond, acts on the spins attached to the vertex $v$. The $B_{p}$ operator,
denoted by the blue diamond, acts on the spins which form the face $p$. The red squares denote $A_{v}$ violations which appear in pairs at the end of the ribbon operator generated by $\sigma_{z}^{i}$. Similarly,
the orange squares denote $B_{p}$ violations, which also appear in pairs at the end of the ribbon operator generated by $\sigma_{x}^{j}$.}
\end{figure}
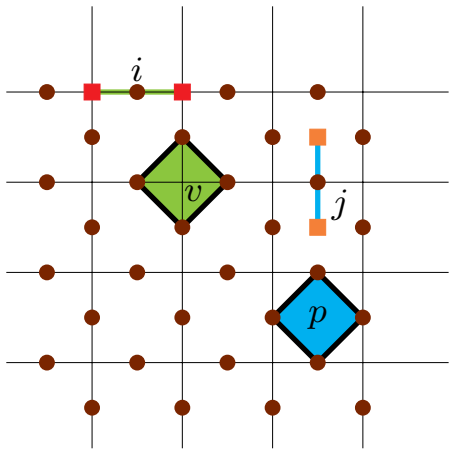

We briefly review the physics behind the toric code model \cite{Kitaev2003}, the simplest case of more general class of models known as quantum double models \cite{Beigi2011}.
Consider a square lattice with vertices denoted by $v$, the faces denoted by $p$ and with spins on edges of the lattice on a torus, as in Fig.~\ref{fig1}. The Hamiltonian of the toric code is given by, 
\be \label{eq1}
H_{toric} = -\sum_{v}A_{v} - \sum_{p}B_{p}
\ee
where
$A_{v} = \prod\limits_i \sigma_{x}^{i}$ and 
$B_{p} = \prod\limits_j \sigma_{z}^{j}.$ 

The ground state, $\ket{\psi_{gs}}$, of the toric code Hamiltonian is a simultaneous ground state of all the $A_{v}$ and $B_{p}$ 
operators, implying the relations
\be\label{gs}
\begin{split}
& A_{v}\ket{\psi_{gs}} = \ket{\psi_{gs}}, \forall v, \\
& B_{p}\ket{\psi_{gs}} = \ket{\psi_{gs}}, \forall p.
\end{split}
\ee

The excitations in the model are given by $A_{v}$ violations and $B_{p}$ violations, which are generated by
$\sigma_{z}$ and $\sigma_{x}$ operators respectively, as shown in Fig.~\ref{fig1}. We note that
the ground state manifold is four-fold degenerate as fusing the excitations along non-trivial topological loops
of the torus projects us back into an orthogonal state that satisfies Eq.~\ref{gs} as well. The properties of the excitations and
the ground state degeneracy are both signatures of topological order in the system. One other key signature of topological order is the so called non-vanishing regularization independent constant term in the entanglement entropy for the ground state known as Topological Entanglement Entropy (TEE). 
For toric code it is given by $\log2$. It has been shown that TEE is related to the quantum dimensions of quasi-particle excitations \cite{Preskill2006,Levin2006}, and hence capture the essence of the topological order. 
In this paper, we will extensively use TEE as a signature to quantify the robustness of topological order and thereby use it to predict the topological phase transitions under perturbations of the toric code in 
an open boundary setting.

\section{Gapped boundaries of the Toric code \label{s3}}

In the above section, we have reviewed the well-known idea of topological order in case of toric code on a torus.  
In this section we present a brief review of various open boundary scenarios in the case of toric code, which have been extensively discussed  in Refs.~\onlinecite{Bravyi2006,Beigi2011,Kitaev2012}.
We specifically review the interfaces with the toric code on one side and the vacuum on the other.

\subsection{Boundary classification and the system Hamiltonian \label{ss3a}}

We begin by reviewing the classification of the boundaries which has been presented in Ref.~\onlinecite{Beigi2011} for 
general quantum double models. Given a quantum double $D(G)$, characterized by a finite group $G$ (abelian or non abelian) the possible boundaries are 
classified by the subgroups $K$ of $G$. In the case of toric code, which is equivalent to $D(Z_{2})$ we look at the 
subgroups of $Z_{2}$ which are given by $\{\{e\}, Z_{2}\}$ where ${e}$ is the identity of the group $Z_{2}$. We will refer to the case where the boundary is decorated 
with $\{e\}$ ($K$=$\{e\}$), as identity as boundary and the case where the boundary is decorated with $Z_{2}$ ($K$=$Z_{2}$), as group as boundary.

\begin{figure}[t!]
\begin{tikzpicture}[scale=0.45]
\draw[line width=0.5pt, fill=LimeGreen](3,8)--(4,9)--(5,8)--(4,7)--(3,8)--cycle;
\draw[line width=0.5pt, fill=ProcessBlue](5,4)--(4,5)--(5,6)--(6,5)--(5,4)--cycle;
\draw[line width=0.5pt, fill=MidnightBlue](1,2)--(2,3)--(1,4)--cycle;
\draw[step=2cm,black,very thin] (0.1,2.) grid (7.9,9.9);
     \foreach \x in {4} 
 	  \foreach \y in {3,5,7,9}
 	      \fill[Brown] (\x,\y) circle (7pt);
     \foreach \x in {3, 5} 
 	  \foreach \y in {2,4,6,8}
	      \fill [Brown] (\x,\y) circle (7pt);
     \foreach \x in {2,6} 
 		\foreach \y in {3,5,7,9}
 	  	    \fill[Brown] (\x,\y) circle (7pt);
     \foreach \x in {1,7} 
 		\foreach \y in {2,4,6,8}
 	  	    \fill[Brown] (\x,\y) circle (7pt);
\draw (0.1,-1.05) -- (3.4,-1.05);
\foreach \x in {2,4,6}
	\draw[black,dashed] (\x.,0) -- (\x.,2.);
\fill[red] (0.1, -.8) rectangle (0.6,-1.3);
\draw (4.6,-1.05) -- (7.9,-1.05);
\fill[red] (7.4, -.8) rectangle (7.9,-1.3);
\node[scale=1.5] at (4,-1) {$L$};
\node[scale=1.] at (4,-3) {(a)};
\draw[-{>[scale=2.0]}] (9,3.75) -- (9,0);
\node[scale=1.5] at (9,4.75) {$R$};
\draw[-{>[scale=2.0]}] (9,6) -- (9,9.9);
\end{tikzpicture}
\hspace{0.25cm}
\begin{tikzpicture}[scale=0.45]
\draw[line width=0.5pt, fill=LimeGreen](4,7)--(3,8)--(4,9)--(5,8)--(4,7)--cycle;
\draw[line width=0.5pt, fill=ProcessBlue](5,4)--(4,5)--(5,6)--(6,5)--(5,4)--cycle;
\draw[line width=0.5pt, fill=ForestGreen](2,3)--(3,4)--(2,5);
\draw[step=2cm,black,very thin] (2,2) grid (6,9.9);
     \foreach \x in {4} 
 		\foreach \y in {3,5,7,9}
 	  	   \fill[Brown] (\x,\y) circle (7pt);
     \foreach \x in {3, 5} 
 		\foreach \y in {2,4,6,8}
 	  	   \fill[Brown] (\x,\y) circle (7pt);
     \foreach \x in {2, 6} 
 		\foreach \y in {3,5,7,9}
 	  	   \fill[Brown] (\x,\y) circle (7pt);
\node[scale=1.5] at (4,-1) {$L$};
\node[scale=1.] at (4,-3) {(b)};
\draw (2.1,-1.05) -- (3.5,-1.05);
\fill[red] (2.1, -.8) rectangle (2.6,-1.3);
\draw (4.6,-1.05) -- (6.,-1.05);
\fill[red] (5.5, -.8) rectangle (6.,-1.3);
\foreach \x in {2,4,6}
	\draw[black,dashed] (\x.,0) -- (\x.,2.);
\end{tikzpicture}
\caption{Toric code with open boundaries on a cylinder with length $L$ and radius $R$. (a) Toric code with identity as boundary. (b) Toric code with group as boundary.
The brown dots represent the physical spins, the light green diamond represents the $A_{v}$ operator, the light
blue diamond represents the $B_{p}$ operator, and the dark blue half diamond represents the modified $B_{p}^{'}$ operator
for identity as boundary in (a), and the dark green half diamond represents modified $A_{v}^{'}$ operator for group as boundary in (b).} 
\label{fig2}
\end{figure}
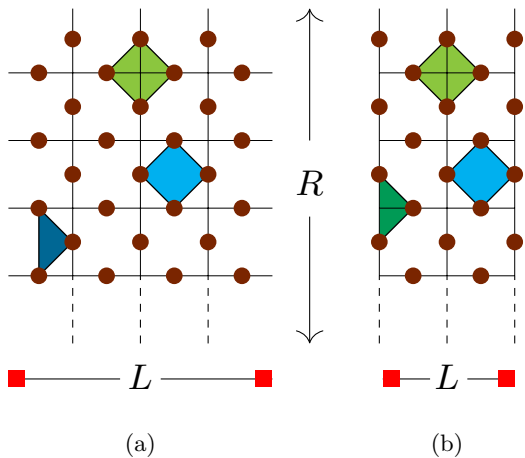

To be explicit, we consider the toric code on a cylinder. We define the Hamiltonian of the system by adding additional boundary terms to the bulk Hamiltonian, the latter being the familiar toric code model. Thus, the Hamiltonian has the form
\be \label{mod1}
 H_{boundary} = H_{toric}^{bulk} - \sum_{i \in boundary}O_{i}.
\ee
For the case of identity as boundary, $K=\{e\}$, we have $O_{i} = B_{p}^{'}$,
where $B_{p}^{'} = \prod_{i=1}^{3}\sigma_{z}^{i}$, with $i$ representing the spins on the boundary faces, as shown in Fig.~\ref{fig2}(a). In the case of group as boundary, $K=Z_2$, we have $O_{i} = A_{v}^{'}$
where $A_{v}^{'} = \prod_{i=1}^{3}\sigma_{x}^{i}$, with $i$ now running over the spins at the boundary vertices, see Fig.~\ref{fig2}(b).
  
Alternatively, boundaries can also be classified by the 
behavior of the excitations at the boundary. For a given quantum double model $D(G)$, the excitations are given by the irreducible representations of
the centralizers of the conjugacy class of $G$. Given a choice of boundary, every excitation
either vanishes and is called a condensing excitation, or it is retained at the boundary, which is known as the non-condensing 
excitation. Ref.~\onlinecite{Beigi2011} provides a mechanism to identify whether an excitation condenses at a particular 
choice of boundary. This condensing/non-condensing behavior of the excitation can be used to classify different boundary conditions. For instance, in the toric code with 
open boundaries, identity as boundary corresponds to $A_{v}$ excitation condensing on the boundary, while group as boundary corresponds to $B_{p}$ excitation 
condensing on the boundary, and vice-versa.

For both the boundary cases, one of the ground states is given by
\be 
\ket{\psi_{gs}}_{boundary} = \mathcal{N}\prod_{v}(1+A_{v})\ket{\textbf{0}},
\ee
where $\ket{\textbf{0}} = \ket{00...0}$, $\ket{0} = \begin{pmatrix} 1 \\ 0\end{pmatrix}$, $\mathcal{N}$ is a normalization constant, and the product over $v$ is modified to include the vertices depending on the boundary.

Our main goal is to study the robustness of topological order in an open boundary setting with respect to 
perturbations generated by magnetic fields applied in different directions. We lay a particular emphasis on 
identifying phase transitions between a topologically ordered phase (for weak perturbations) and a topologically 
trivial phase (for strong perturbations). We use different methods to identify various signatures of topological 
order to get a consistent picture. This way, our results also serve as a benchmark for the different methods to 
classify the topological properties of a system.

In our analysis, we are interested in the 
topological properties of an infinitely large system. While there 
are different ways how to approach the thermodynamic limit in the 
cylindrical geometry introduced in the previous section, we are 
interested in the case where the boundary contribution is extensive. Hence, we will 
focus on a quasi-1D geometry where the length $L$ of the cylinder is 
finite, while the radius $R$ diverges.

\subsection{Perturbation in the presence of boundaries}
We begin by studying the effect of perturbations of the form $\sum_{i}\sigma_{z}^{i}$, $\sum_{i}\sigma_{x}^{i}$ on the Hamiltonians generated by Eq.\hspace{1mm}\ref{mod1}. In particular, we will investigate cases where 
one of two perturbations is present at a given time.
We could have simultaneously turned both of them at the same time to get the full phase diagram, as in toric code on a torus \cite{Vidal2009, Nikolay2012}, but we leave this investigation to the future.

\subsubsection{Exact Hamiltonians}
Given that we have two different boundary conditions and we consider two different types of perturbation, $\sum_{i}\sigma_{x}^{i}$ and $\sum_{j}\sigma_{z}^{j}$, we have the following cases:
\begin{enumerate} 
\item  In the first case, the perturbed Hamiltonian has the form
    \be \label{EH_idpx}
      H_{idpx(grpz)} =  H_{id(gr)} - h_{x(z)}\sum_{i}\sigma_{x(z)}^{i}.
    \ee
    Here, the perturbation commutes with the $A_{v}$ $(B_{p})$ operator, and anti-commutes with the $B_{p}$ $(A_{v})$ operator leading only to the $B_{p}$ $(A_{v})$ violations. 
    As we have seen in Sec.~\ref{ss3a}, we know that $B_{p}$ $(A_{v})$ violations are contained in the identity (group) as boundary case. 
      As the above violations are both non-condensing, they will give rise to the same effective non-condensing Ising Hamiltonian, which we shall discuss later in detail.

\item In the second case, the perturbed Hamiltonian is given by
    \be \label{EH_grpx}
        H_{idpz(grpx)} = H_{id(gr)} - h_{z(x)}\sum_{i}\sigma_{z(x)}^{i}.
    \ee
    In this scenario, the perturbation commutes with the $B_{p}$ $(A_{v})$ operator, and anti-commutes with the $A_{v}$ $(B_{p})$ operator leading only to $A_{v}$ $(B_{p})$ violations. 
    Again, from the arguments laid out in Sec.~\ref{ss3a}, we know that $A_{v}$ $(B_{p})$ violations
    condense in identity (group) as boundary case.  As these violations are both condensing, they will 
    both lead us to the same effective condensing Ising Hamiltonian.
\end{enumerate}

\subsubsection{Ising Hamiltonians}

In the following, we map the above exact Hamiltonians to associated Ising Hamiltonians. We briefly motivate the map 
and leave the detailed explanation to Appendix~\ref{appA}. We consider a situation where the energy cost of one 
type of excitation (i.e., either $A_v$ or $B_p$) is set to infinity and hence, only the other type of excitation can
be observed in the system \cite{Vidal2011}. As in Ref.~\onlinecite{Vidal2011}, we construct an equivalent model, 
where the degrees of freedom are the excitations of the system and not the individual spins. As any vertex or 
plaquette can contain only one or zero excitations, it is natural to use Ising spin $1/2$ variables to describe 
the effective model. In the bulk, each term of the pertubation creates two excitations on adjacent vertices 
(plaquettes), see Fig.~\ref{fig1}, corresponding to an effective Ising interaction.

However, at the boundary the interaction is captured by the way the excitation behaves at the particular choice of 
boundary. For the case where excitations condense at the boundary, we can always create single excitations in the 
bulk, implying the spins in the excitation picture near the boundary can be flipped independently, resulting in 
the Hamiltonian 
\be \label{CI} 
H_{ci} = -h\bigg(\sum_{i,j}\mu_{i}^{x}\mu_{j}^{x} + \sum_{k \in boundary}\mu_{k}^{x}\bigg) - \sum_{i}\mu_{i}^{z}
\ee
where $H_{ci}$ refers to the Condensing Ising (CI) Hamiltonian with perturbation strength $h$. For the case where
excitations do not condense on the boundary, the nearest neighbor Ising interaction between the excitations at 
the boundary is still intact, resulting in the Hamiltonian
\be \label{NCI}
H_{nci} = -h\sum_{i,j}\mu_{i}^{x}\mu_{j}^{x} - \sum_{i}\mu_{i}^{z}, 
\ee
where $H_{nci}$ refers to the Non Condensing Ising (NCI) Hamiltonian with the perturbation strength, $h$ on the 
Ising interaction.

We verify the above intuitive picture by using the Controlled-NOT
(CNOT) mechanism as in Ref.~\onlinecite{Vidal2011}, which is presented
in Appendix~\ref{appB}. By performing the CNOT mechanism in the open boundary
context, we observe that in addition to the Ising like interactions we
have a vacancy and topological spin (non-local spin-$\frac{1}{2}$
associated to the non-trivial [non-local] loops which project into
different ground states) as observed in the periodic boundary case
\cite{Vidal2011, Sachdev2016}.  Let us summarize the main results from
Appendix~\ref{appB}, focusing on the periodic boundary of the cylinder in the
context of group as boundary under $\sum_{i}\sigma_{x}^{i}$
perturbation (\ref{EH_grpx}).  Using the CNOT map, we see that the
interaction at the periodic boundary is captured by
$-h\sum_{p',q'}\mu_{p'}^{x}\otimes L_{x} \otimes \mu_{q'}^{x}$, where
$p'$, $q'$ are Ising spins on either side of the boundary with $L_{x}$
given by \be L_{x(z)} = \prod_{j\in L}\sigma_{x(z)}^{j}, \ee $L$ being
the shorter width of the cylinder.  Similary, we can extend the result
to other Hamiltonians ~\ref{EH_idpx}, ~\ref{EH_grpx} 
where the boundary is coupled by the non-trivial loop operator $L_{x(z)}$ as
above. While the effect of these additional terms is irrelevant for
the thermodynamic properties of the system, they are essential to
capture the relation between a thermodymaic phase transition in the
effective model and the associated breakdown of topological order.

\section{Characterizing the phase transition: Ising map and magnetization \label{s4}}

In this section we analyze the two Hamiltonians, (\ref{CI}) and (\ref{NCI}), to gain an understanding of the 
phase transition in detail. We observe the behavior of magnetization, which acts as the order parameter, with 
respect to the strength parameter $h$. 

\subsection{Non Condensing case}

We begin by studying the NCI Hamiltonian at different perturbation strengths. We note that as perturbation 
strength is increased, long ribbon operators (long strings with excitations at their ends) are penalized and 
shorter loops gain prominence which are captured by Ising interaction. Therefore we expect to capture a 
paramagnetic to a ferromagnetic transition and to gain an insight into the phase transition, we calculate 
the absolute of magnetization $m$, which is given by 
\be
 m =\frac{1}{N}\sum_{i=1}^N\mu_{i}^{x},
\ee
where $N$ is the total number of spins.  We denote the susceptibility  by $\chi$, which is given by
\be
 \chi=
\frac{\partial m}{\partial h}
\ee

\begin{figure}[t!]
\begin{center}
\includegraphics[width=\columnwidth]{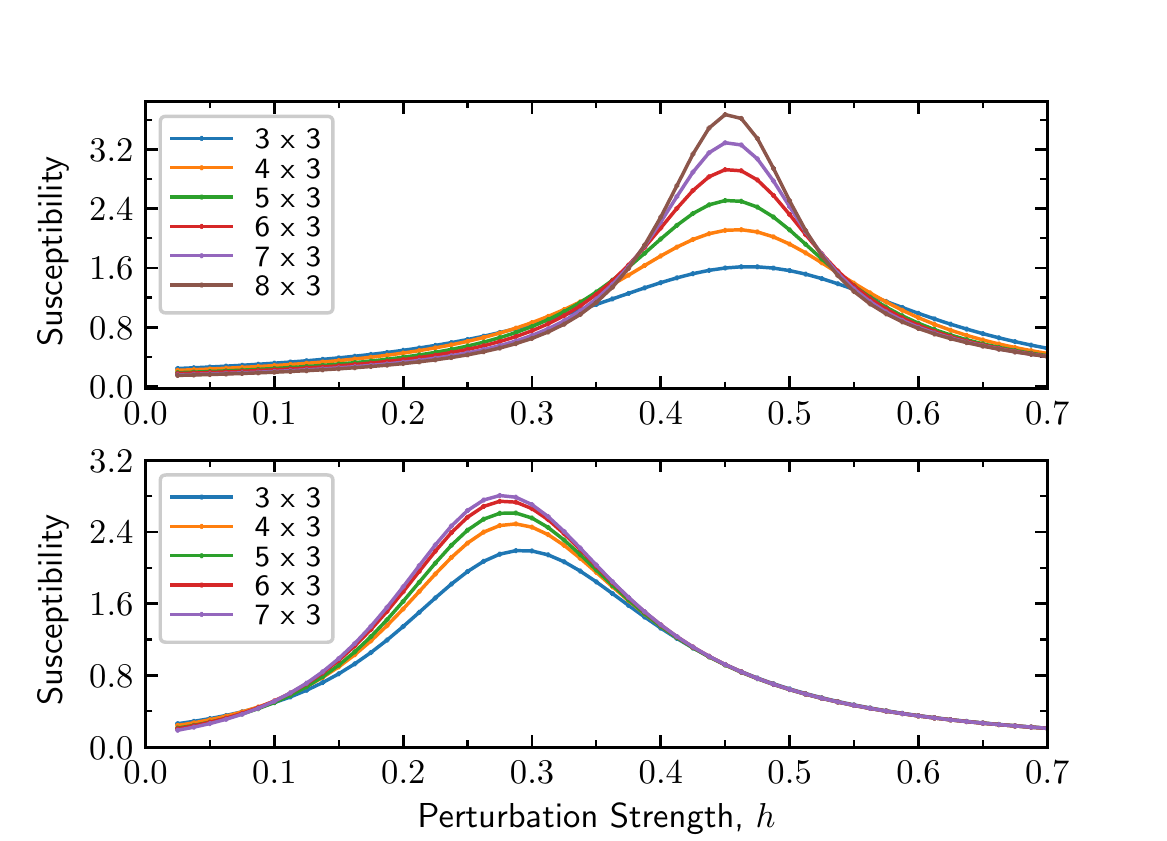}
\end{center}
\caption{Susceptibility of the Ising model equivalent to the non-condensing case (Top)  condensing case (Bottom) }
\label{fig4}
\end{figure}

From Fig.~\ref{fig4}, we infer that there is a phase transition occurring 
at some point between 0.4 and 0.5 as the susceptibility diverges with $R$. To precisely capture the transition 
we perform a finite size scaling analysis\cite{Cardy1996}.

\subsubsection*{Finite Size scaling}
The expression used for finite size scaling is given by

\be
m = N^{-\beta/\nu}\tilde{f}\left([h-h_c]N^{1/\nu}\right), 
\ee
where $h_{c}$ is the critical strength, $m$ is the magnetization, $N$ is the system size, $\nu, \beta$ are the critical 
exponents, $\tilde{f}$ is a scaling function chosen as a degree 6 polynomial. Here we know $m$, $g$, $N$ and fit the data to determine
the coefficients of the polynomial, $\tilde{m}$ and $g_{c}$.  
From the fit, We infer that that the critical strength is given by $h_{c}$ = 0.453$\pm$0.001 and the critical exponents are given by, $\beta$ = 0.1$\pm$0.007, $\nu$=1.131$\pm$0.013
which are in good agreement with the critical exponents of the 2D Ising universality class ($\beta$=0.125, $\nu$=1).

We note that topological order under perturbation 
is more robust in the non-condensing scenario in comparison to the 
periodic boundary case, where the critical strength occurs at 
$h_{c}^{periodic}$=0.328\cite{Nikolay2012}. This surprising result 
can be attributed to the role of quantum fluctuations in both 
cases. Crucially, the topologically ordered phase corresponds to the 
disordered phase of the Ising model (i.e., the paramagnet), and vice 
versa. In our cylindrical setup representing a quasi-1D system, the role 
of quantum fluctuations is stronger than on 2D setup on a torus. Hence, in 
the context of constructing a robust memory, it appears that is more beneficial 
to store a single qubit in a quasi-1D setup rather than two qubits on a torus.

\subsection{Condensing case}

We analyze the condensing case analogously to the non-condensing case. Observing the CI Hamiltonian (\ref{CI}), in the thermodynamic limit, we see that $Z_{2}$ symmetry of the Ising model is broken
as soon as the perturbation is turned on. This is because of the localized excitations which appear at vertices/faces in bulk which share an edge with the boundary,
Also, the topological coupling terms provide further insight into the breaking of ground 
state degeneracy, which is a signature of topological order. 

In the presence of the perturbation, the effective Ising Hamiltonian including the topological coupling term is 
given by (cf. App.~\ref{appB})
\be\label{CIEff} 
\begin{split}
&H_{ci}^{eff} =   - \sum_{i}\mu_{i}^{z} -h\bigg(\sum_{i,j}\mu_{i}^{x}\mu_{j}^{x} + \sum_{k \in boundary}\mu_{k}^{x}\\
& \hspace{2cm} + \sum_{(p,q)}\mu_{p}^{x}\otimes L_{x} \otimes \mu_{q}^{x}\bigg).
\end{split}
\ee

We now perform a mean-field decoupling for the Ising model and look 
into the resulting coupling to the topogical sector. Then, we obtain 
an effective topological coupling of the form $h_{eff}^{MF} = 
h\braket{\mu_{p}^{x}}\braket{\mu_{q}^{x}} L_{x}$. In the condensing 
case, the presence of the $Z_2$ breaking terms lead to 
$\braket{\mu_{p}^{x}}\braket{\mu_{q}^{x}} \neq 0$ for any finite 
$h$. Hence, the topological degeneracy is lifted once the 
perturbation is turned on and topological order is destroyed. On the 
other hand, in the non-codensing case, we have 
$\braket{\mu_{p}^{x}}\braket{\mu_{q}^{x}} = 0$ for 
$h<h_c$. Consequently, the topological sector is decoupled from the 
Ising model and topological order remains intact. Strikingly, in the 
ferromagnetic phase, $h>h_c$, the expectation values become finite 
since they represent the order parameter of the ferromagnet. Again, 
this leads to a lifting of the topological degeneracy and a 
breakdown of topological order. Including the topological terms in 
the CNOT map therefore provides a deep insight into the connection 
between the thermodynamic transition in the effective Ising model 
and the associated breakdown of topological order.

As in the non-condensing case, we compute the magnetization to numerically verify the above claim. From Fig.~\ref{fig4}, we observe that there is 
no divergence in the susceptibility  with increase in the perturbation strength. Therefore, to further strengthen and numerically verify the above claim we revert back to the original model and study
various signatures of topological order. As we have already established the non-condensing scenario, we use the topological signatures in the non-condensing case to benchmark our analysis of the condensing case.

\section{Closer look at the Condensing case \label{s5}}

We analyze the following signatures to gain an insight into the robustness of topological order:

\begin{enumerate}
 \item Breaking of the ground state degeneracy
 \item Topological Entanglement Entropy
 \item Minimally Entangled States (Two minima in topologically ordered phase to one in trivial phase)
\end{enumerate}

\subsection{Energy scaling}

\subsubsection{Identity as boundary}
\begin{figure}[t!]
\begin{center}
\includegraphics[width=\linewidth]{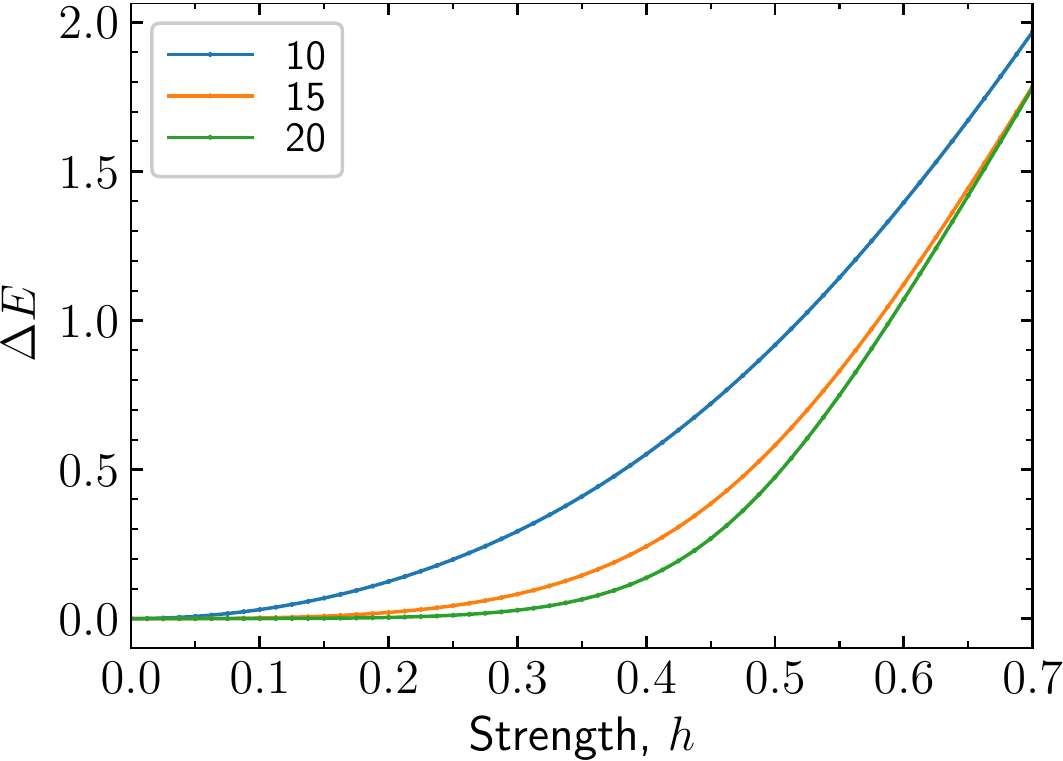}
\end{center}
\caption{Energy Difference, $\Delta E$, in the presence of perturbation for the non-condensing case (identity as boundary).}
\label{fig7}
\end{figure}

\begin{figure}
   \begin{overpic}[width=\linewidth]{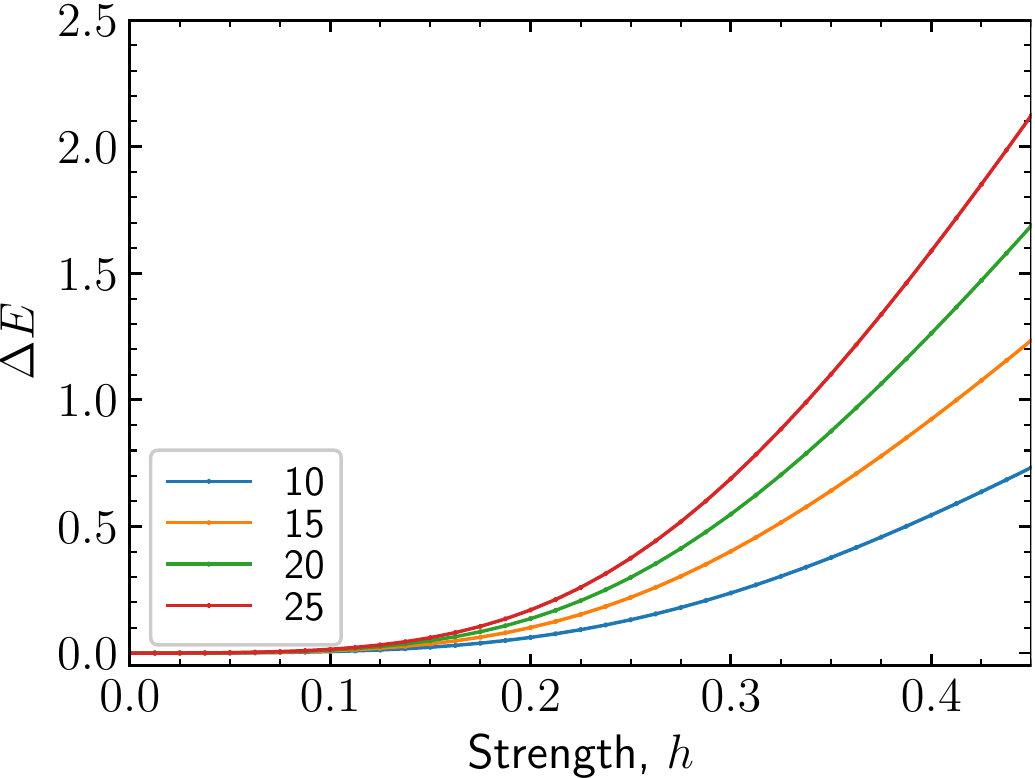}
     \put(19,36){\includegraphics[width=0.575\linewidth]{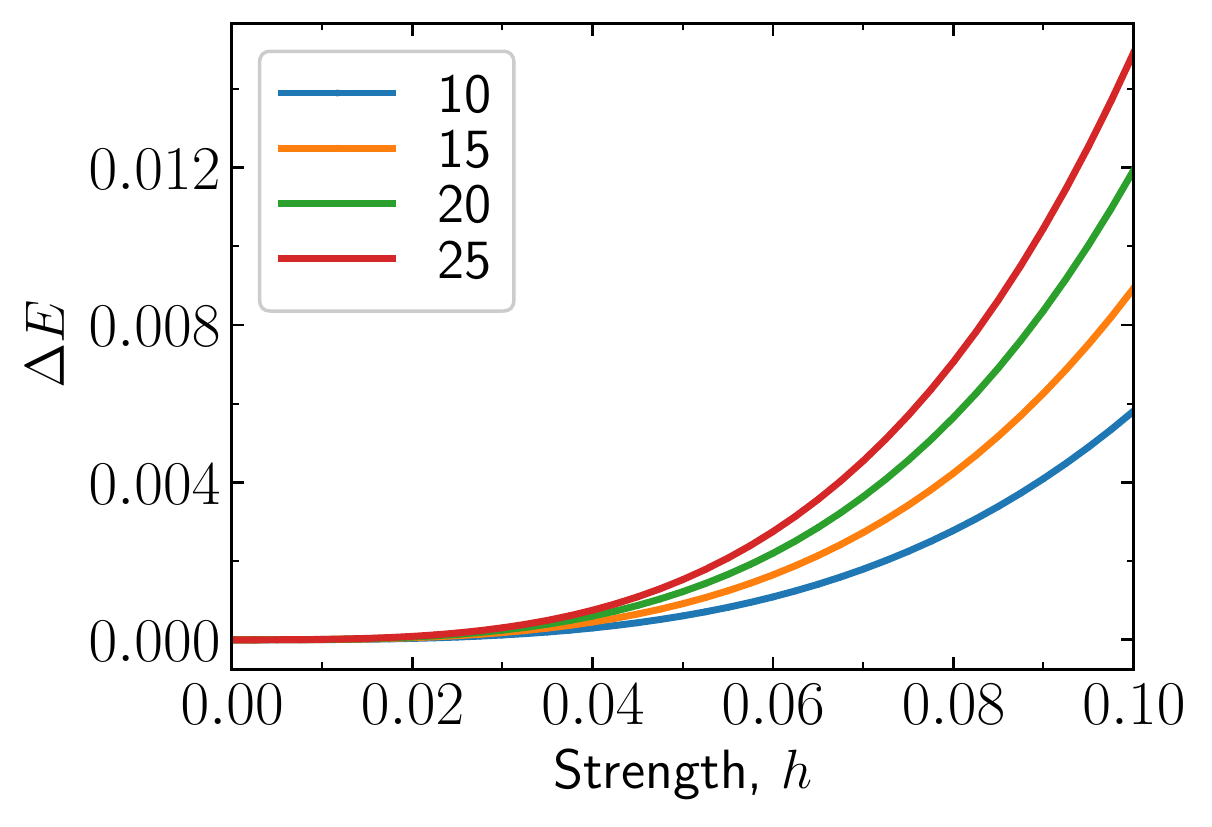}}
  \end{overpic}
\caption{Energy Difference, $\Delta E$, in the presence of perturbation for the condensing case with identity as boundary. (Inset) $\Delta E$ for strength range much closer to zero.}
\label{fig8}
\end{figure}

We begin by analyzing the ground state degeneracy of the Hamiltonian $H_{idpz}$, which we know has two degenerate 
ground states \cite{Juven2015} at $h_{x}=0$. In the thermodynamic limit, we expect the ground state degeneracy to break as soon as we turn on the perturbation, which we aim 
to observe in terms of the energy difference ($\Delta E$). Using the fact that there is a phase transition in the non-condensing case, we compare Fig.~\ref{fig7} and Fig.~\ref{fig8}, which 
depict the behavior of $\Delta E$ to gain an insight into the understanding of the condensing case. We see that in the non-condensing case as we approach the thermodynamic limit, there is a suppression
in $\Delta E$ for $h < h_{c}$, which is as expected \cite{Nayak2007}, but in the condensing case, $\Delta E$ increases with perturbation strength as well with an increase in system size, with no 
suppression. Extrapolating the results to the thermodynamic limit, we can conclude that energy gap opens as soon as the perturbation is turned on.

\subsubsection{Group as boundary}
\begin{figure}[t!]
\begin{center}
\includegraphics[width=\linewidth]{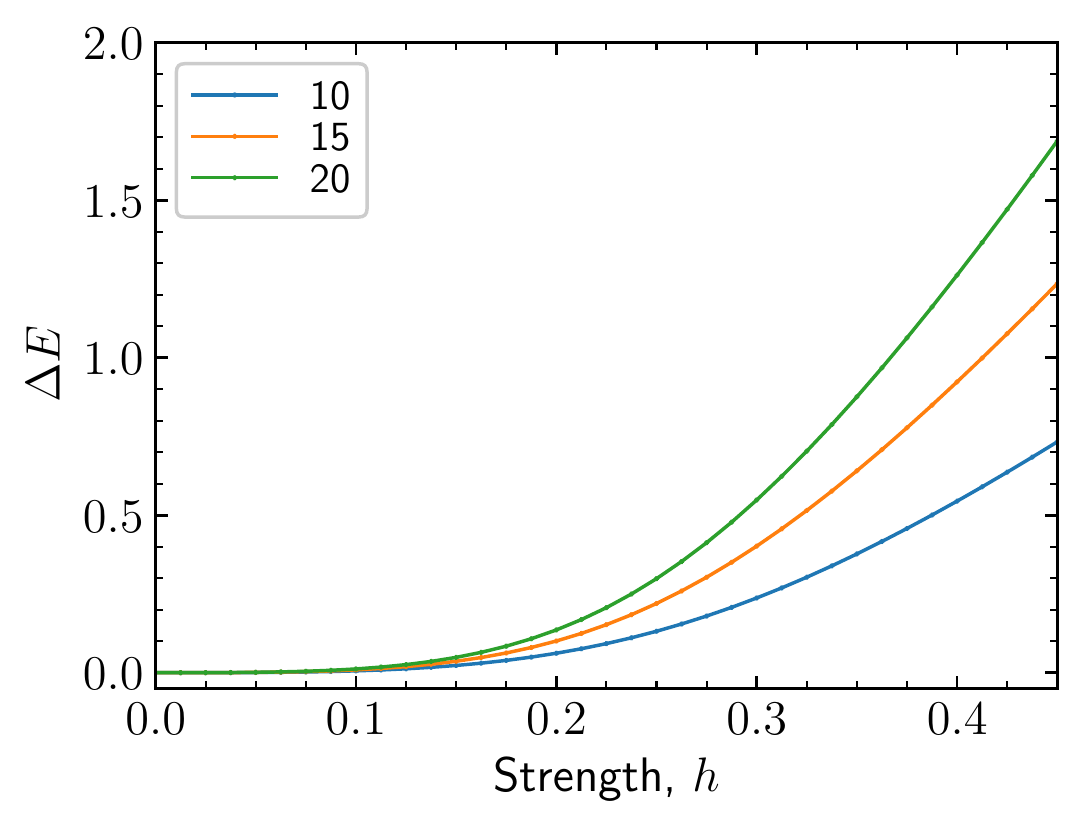}
\end{center}
\caption{Energy Difference, $\Delta E$, in the presence of perturbation for the condensing case with group as boundary, we observe that the nature of the graph is similar to the case of 
identity as boundary as in Fig.~\ref{fig8}.}
\label{fig9}
\end{figure}

As a consistency check, we consider the Hamiltonian $H_{grpx}$ and compute $\Delta E$, as expected 
the degeneracy is lifted as soon as the perturbation is turned on as in Fig.~\ref{fig9}. Therefore, for the condensing class we conclude that 
$\Delta E > 0$ as soon as the perturbation is turned on, which strengthens the claim that for $h_{x} >0$ the phase is broken, making it topologically trivial.  

\subsection{Topological Entanglement Entropy}
One other key signature of topological order is the topological term in the entanglement entropy \cite{Levin2006,Preskill2006}. Consider a region, say $A$, on the lattice, and denote the reduced density matrix by $\rho_{A}$. The 
von-Neumann entropy $S_{A}$ given by $S_{A}$ = -tr($\rho_{A}$ln$\rho_{A}$) scales as following 
\be
S_{A} = a\,L_{cut} - \gamma
\label{ent}
\ee
where $L_{cut}$ is the length of the cut and $\gamma$ is identified as the topological term and is called Topological Entanglement Entropy (TEE) which is a signature of topological order.

\begin{figure}[t!]
\begin{tikzpicture}[scale=0.45]
	\draw[step=2.cm,black,very thin] (0.1,2.) grid (7.9,7.9);
     \foreach \x in {2,4,6} 
 		\foreach \y in {3,5,7}
 	  	   \fill[Blue] (\x,\y) circle (6pt);
     \foreach \x in {1,3,5,7} 
 		\foreach \y in {2,4,6}
           \fill [Blue] (\x,\y) circle (6pt);
\draw[line width=2pt, red](6.5,0.1)--(6.5,7.9);
\node[scale=1.] at (4,-1) {(a)};
\foreach \x in {2,4,6}
	\draw[black,dashed] (\x.,0) -- (\x.,2.);
\end{tikzpicture}
\hspace{0.2cm}
\begin{tikzpicture}[scale=0.45]
	\draw[step=2.cm,black,very thin] (2.,2.) grid (8.,7.9);
     \foreach \x in {2,4,6,8} 
 		\foreach \y in {3,5,7}
 	  	   \fill[Blue] (\x,\y) circle (6pt);
     \foreach \x in {3,5,7} 
 		\foreach \y in {2,4,6}
 	  	   \fill[Blue] (\x,\y) circle (6pt);
\draw[line width=2pt, red](2.5,0.1)--(2.5,7.9);
\node[scale=1.] at (5,-1) {(b)};
\foreach \x in {2,4,6,8}
	\draw[black,dashed] (\x.,0) -- (\x.,2.);
\end{tikzpicture}
\caption{Different cuts used for the computation of entropy. \newline (a) Cut for identity as boundary (b) Cut for group as boundary. The region used for the computation of entropy always includes the
boundary and it can be either the region to the left or right of the cut. For computational purposes, we choose the region to the right of cut in (a) and left of the cut in (b) for computing the entropy.}
\label{fig13}
\end{figure}
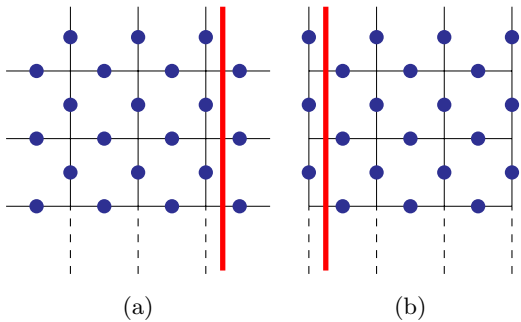

To compute $\gamma$, we consider a topologically non-trivial cut which winds around the surface of the cylinder as in the Fig.~\ref{fig13} and replicate the method used in Ref.~\onlinecite{Balents2012}. 
Note that the cut that we are employing is different than the ones used in Refs.~\onlinecite{Chen2018,Cheipesh2018}, as our choice contains only a single boundary.
The length of the region $A$, from which we extract the TEE, scales with $R$ and the width of the region $A$ is fixed. Therefore, in our case the above equation (\ref{ent}) changes to
\be
S_{A} = a\,R - \gamma
\ee
consequently computing the entropy for different $R$ and fitting $S_{A}$ versus $R$ gives us $\gamma$, which is the y-intercept of the fit.

\subsubsection{Non-Condensing case}
\begin{figure}[t!]
\begin{center}
\includegraphics[width=0.9\linewidth]{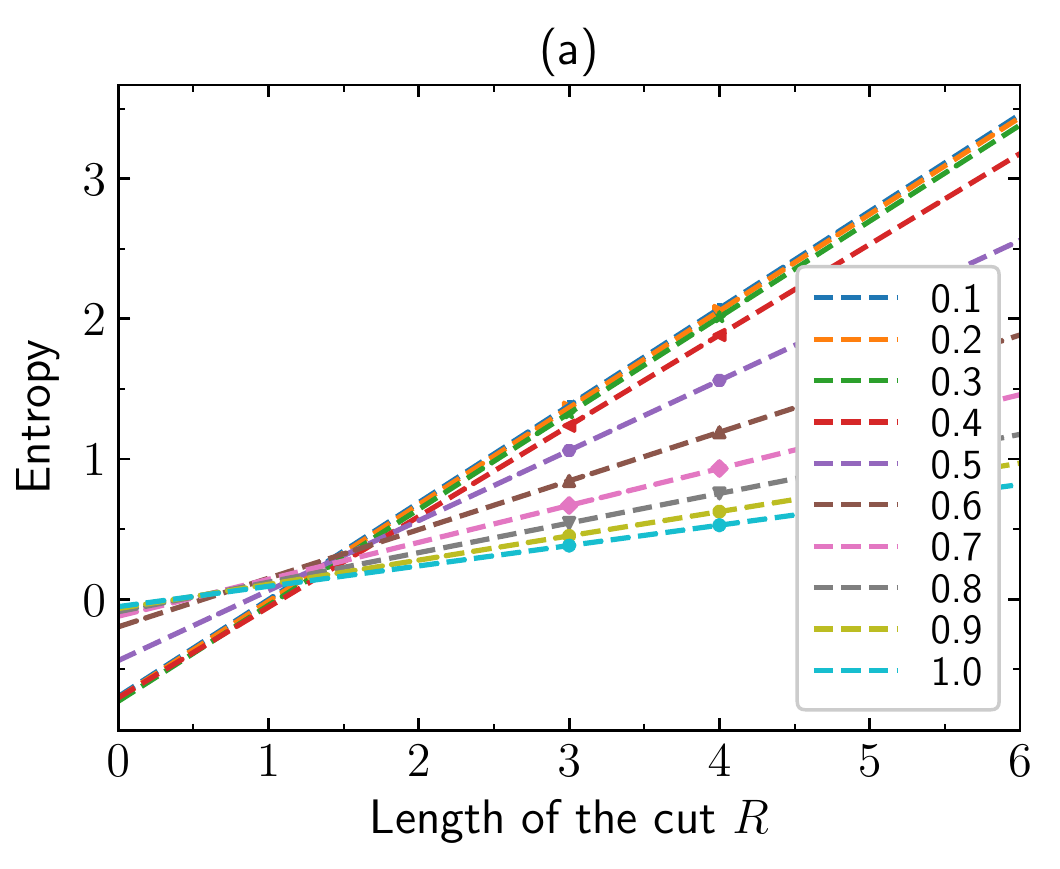}
\includegraphics[width=0.9\linewidth]{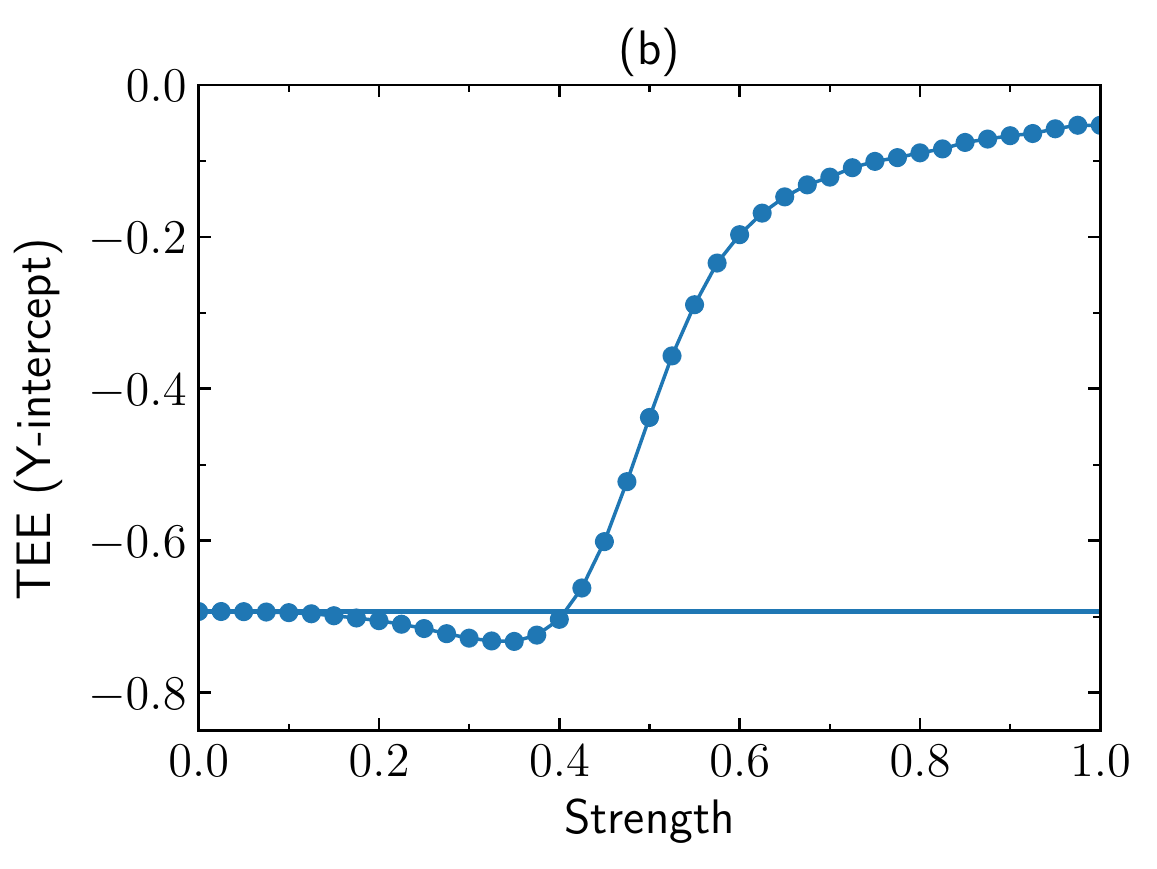}
\end{center}
\caption{Extracting TEE in the non-condensing case. (a) For a given strength, we consider cuts of $R$ = 3, 4 for computing the entropy, $S_{A}$, where region $A$ is as in Fig.~\ref{fig13} and 
then fit $S_{A}$ versus $R$ to extract the TEE,  which is the y-intercept of the fit.
(b) TEE at different perturbation strengths. As the perturbation strength is increased, TEE scales from $\gamma$ = $\log2$ to $\gamma$=0, signaling a phase transition.}
\label{fig14}
\end{figure}

We first verify the already established fact of phase transition for the non-condensing class using TEE. Consider the non trivial cut as in 
the Fig.~\ref{fig13}, either of the boundary conditions can be considered with a suitable perturbation that results in a non-condensing scenario.
Here we consider the Hamiltonian $H_{idpx}$ and extract the TEE using Fig.~\ref{fig14}(a), which is the y-intercept of the 
plot between $S_{A}$ and $R$. We plot the TEE against the strength to identify the transition point as in the Fig.~\ref{fig14}(b). 
From Fig.~\ref{fig14}(b), we re-establish the fact that in the non-condensing case there is a phase transition from a ordered phase to a trivial phase as TEE scales from $\log2$ 
to 0 with an increase in the perturbation strength. We also note that, numerically the critical strength from the TEE calculation is comparable to the exact results from the magnetization results. 
While we observe a dip in the TEE below $-\log(2)$, we attribute this to a finite size effect since it occurs close to the transition point, where finite size effects are particularly strong.

\subsubsection{Condensing case}
\subsubsection*{a. Identity as boundary}
\begin{figure}[t]
   \begin{overpic}[width=\linewidth]{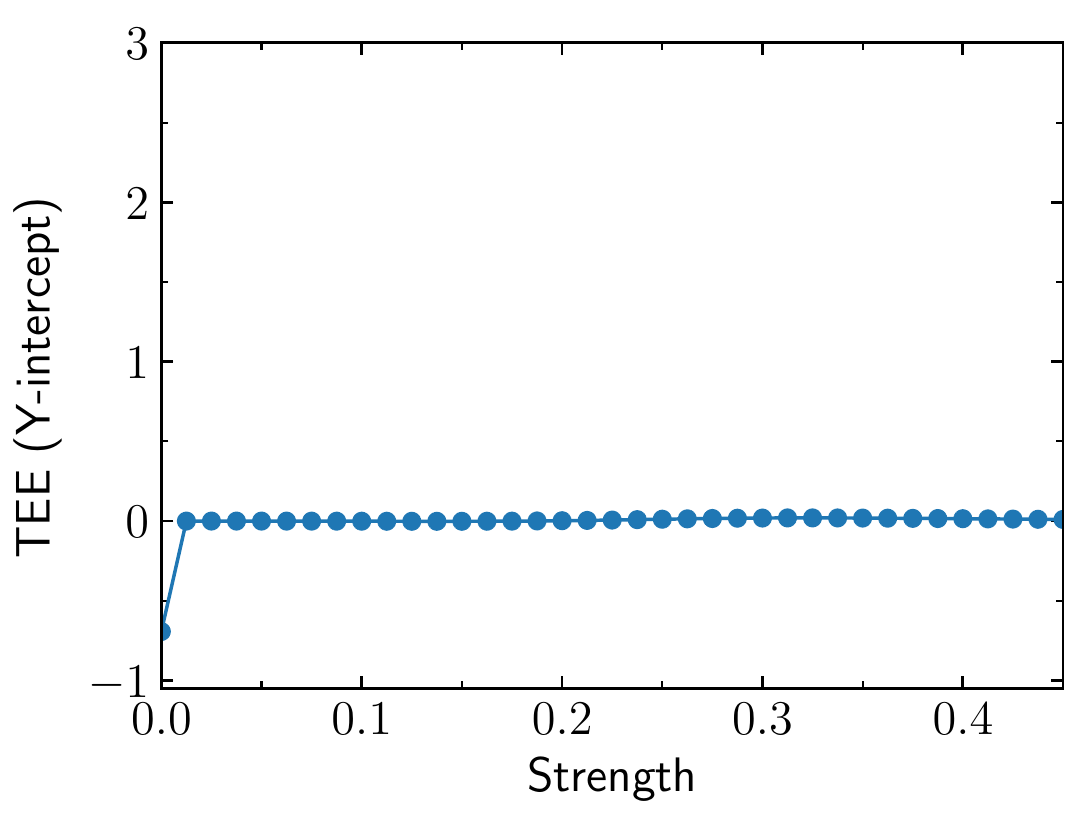}
     \put(35, 30){\includegraphics[width=0.6\linewidth]{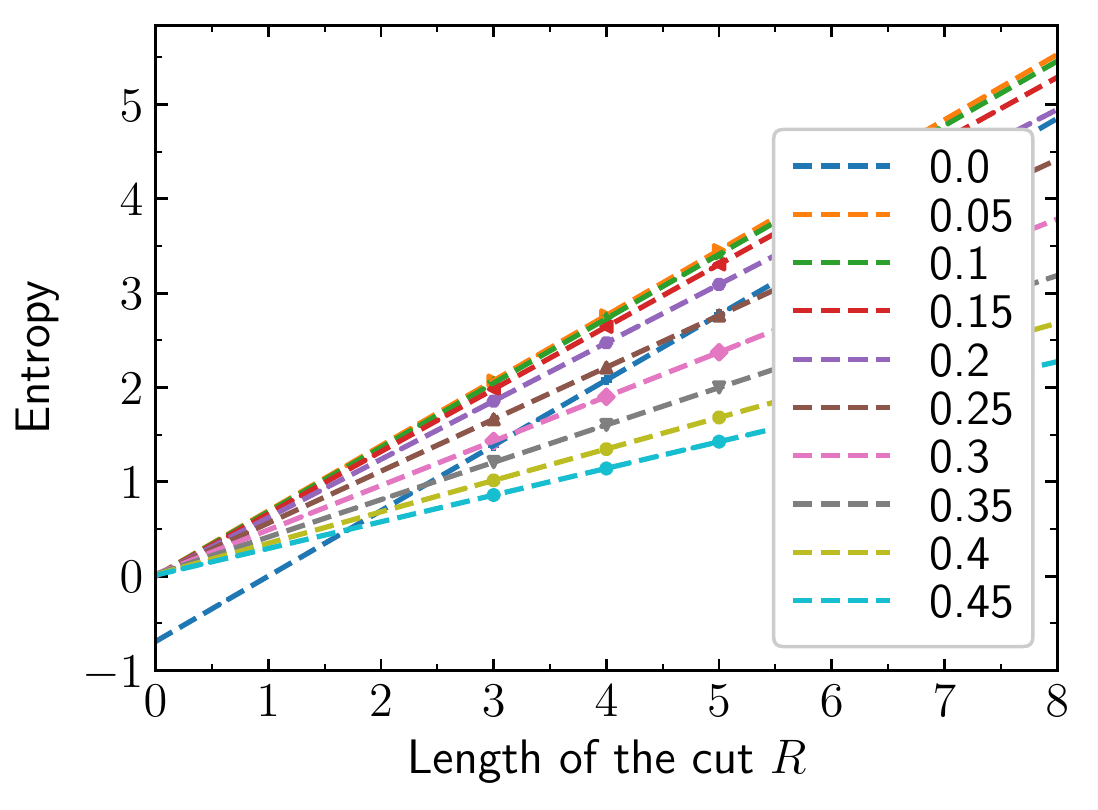}}
  \end{overpic}
\caption{Extracting TEE in the condensing case, with identity as boundary. TEE remains zero as the strength is varied, only for $h=0$ it is $\gamma = \log2$.
(Inset) Entropy versus the length of the cut (a non trivial cut equal to the radius of the cylinder), to extract the TEE (y-intercept) at each strength.}
\label{fig15}
\end{figure}

From the above case, it is clear that we can predict the presence of a transition point by observing the behavior of TEE. For the condensing case, in the earlier sections we 
argued that the phase is trivial as soon as the perturbation is turned on. To further consolidate the claim, we study the TEE behavior in the presence of perturbation for the condensing class. First we consider 
the Hamiltonian $H_{idpz}$ and verify the results for the Hamiltonian $H_{grpx}$. As in the non condensing case, we consider a non-trivial cut along the surface of the cylinder, compute the entropy
and extract the TEE for different perturbation strength. From Fig.~\ref{fig15}, we observe that as soon as perturbation strength is turned on, TEE drops to zero thereby confirming the above mentioned picture.

\subsubsection*{b. Group as boundary}
\begin{figure}[t]
   \begin{overpic}[width=\linewidth]{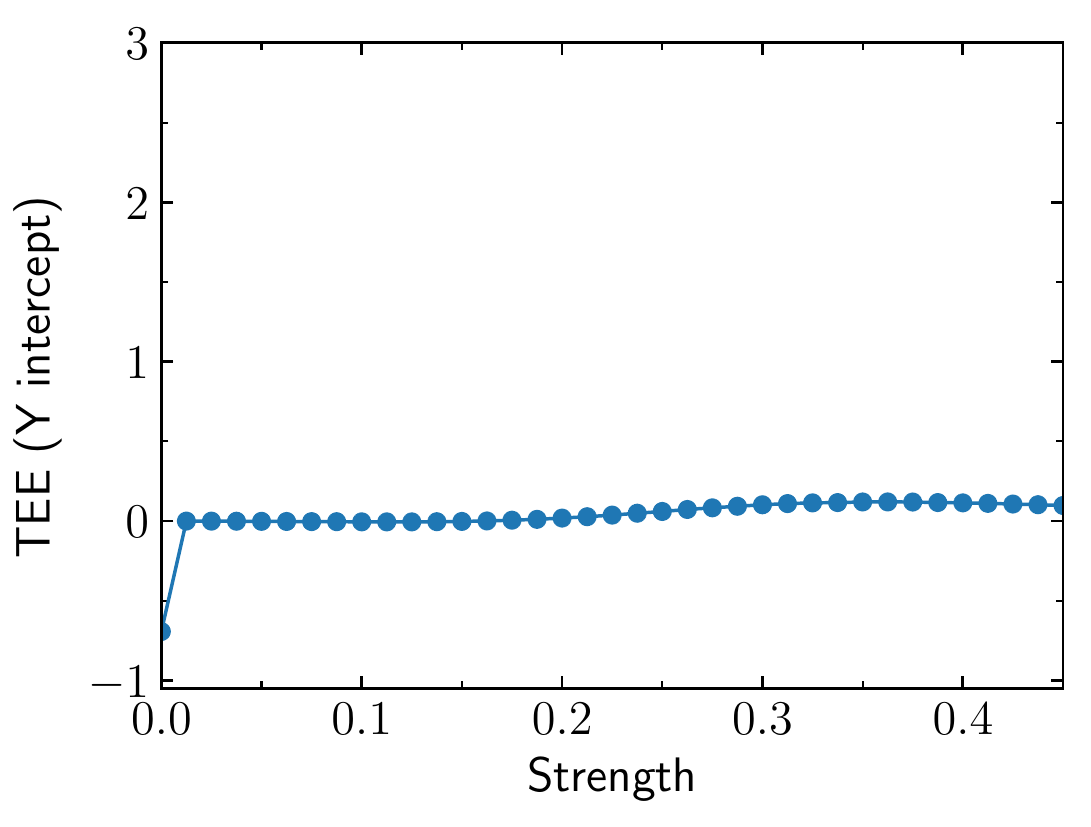}
     \put(35, 32.5){\includegraphics[width=0.6\linewidth]{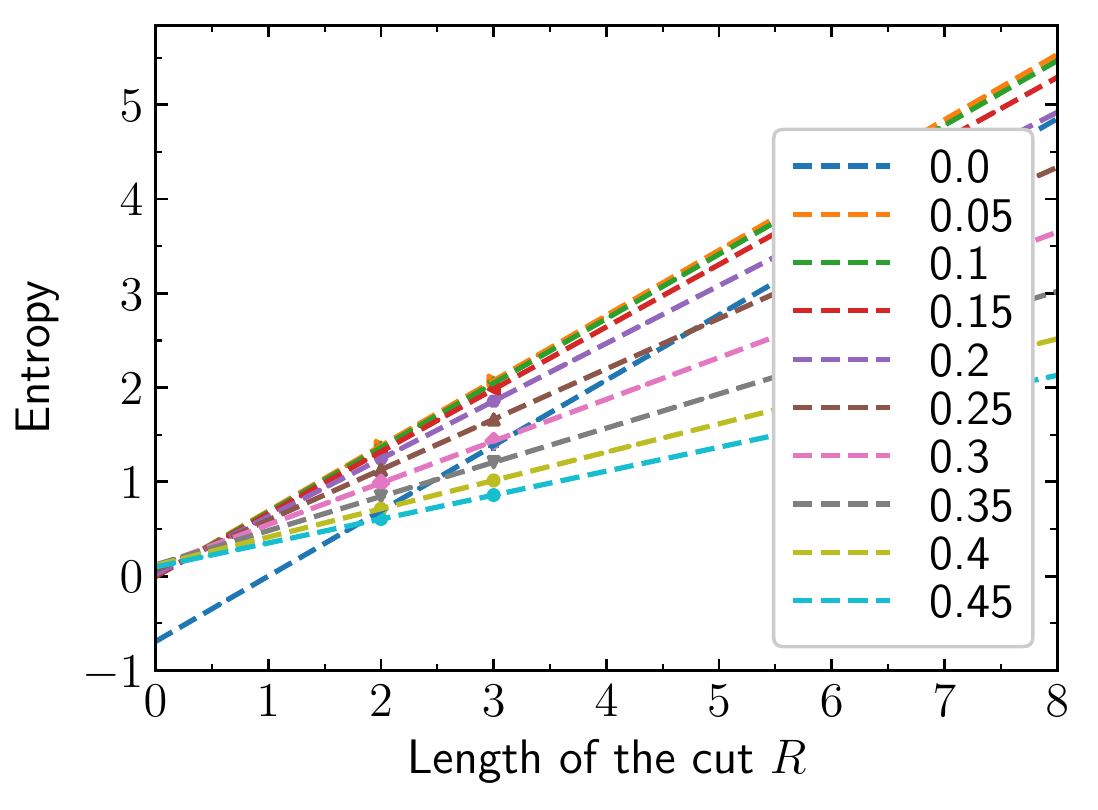}}
  \end{overpic}		
\caption{Extracting TEE in the condensing case, with group as boundary. The nature of TEE remains same as in the identity as boundary case.}
\label{fig16}
\end{figure}

We verify the conclusion from the above section by repeating the process of computing TEE for the condensing case with group as boundary.
From Fig.~\ref{fig16}, we observe that the behavior of TEE is same as above, which supports that for the condensing class, the system 
is topologically ordered at $h=0$ and is trivial for $h>0$.

\subsection{Minimally Entangled States}

\begin{figure*}[t]
\begin{center} 
\includegraphics[width=0.25\linewidth] {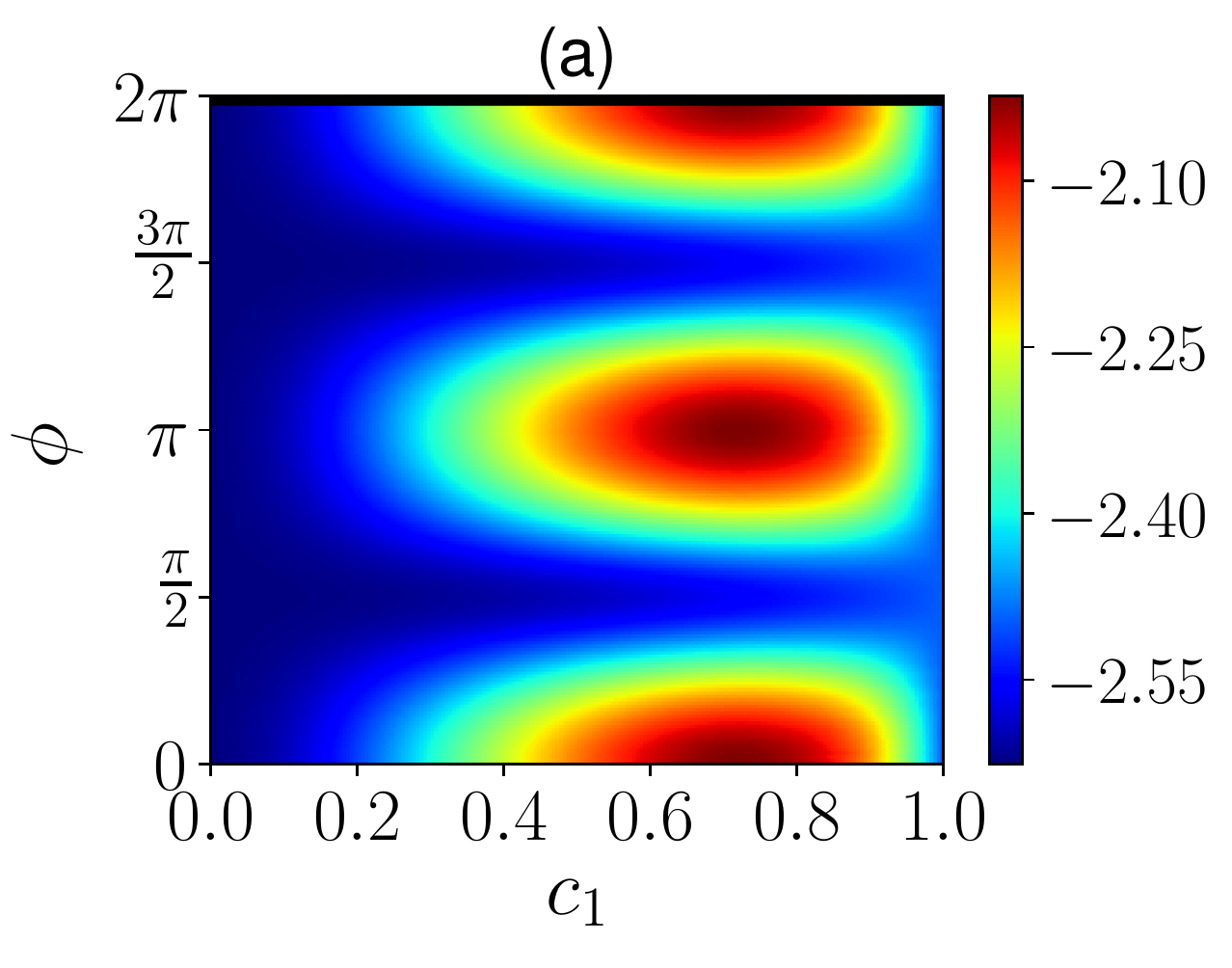}\includegraphics[width=0.25\linewidth]{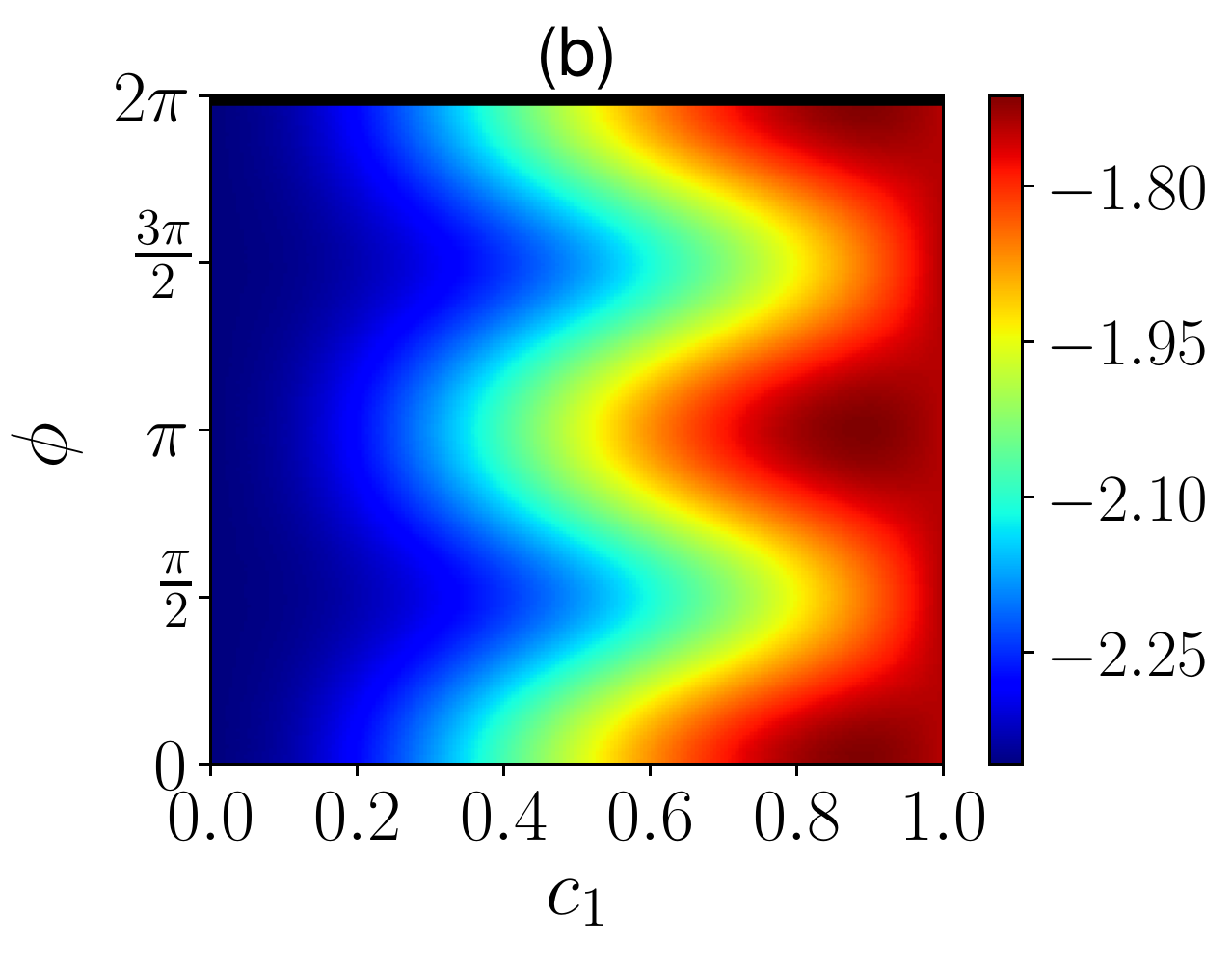}\includegraphics[width=0.25\linewidth]{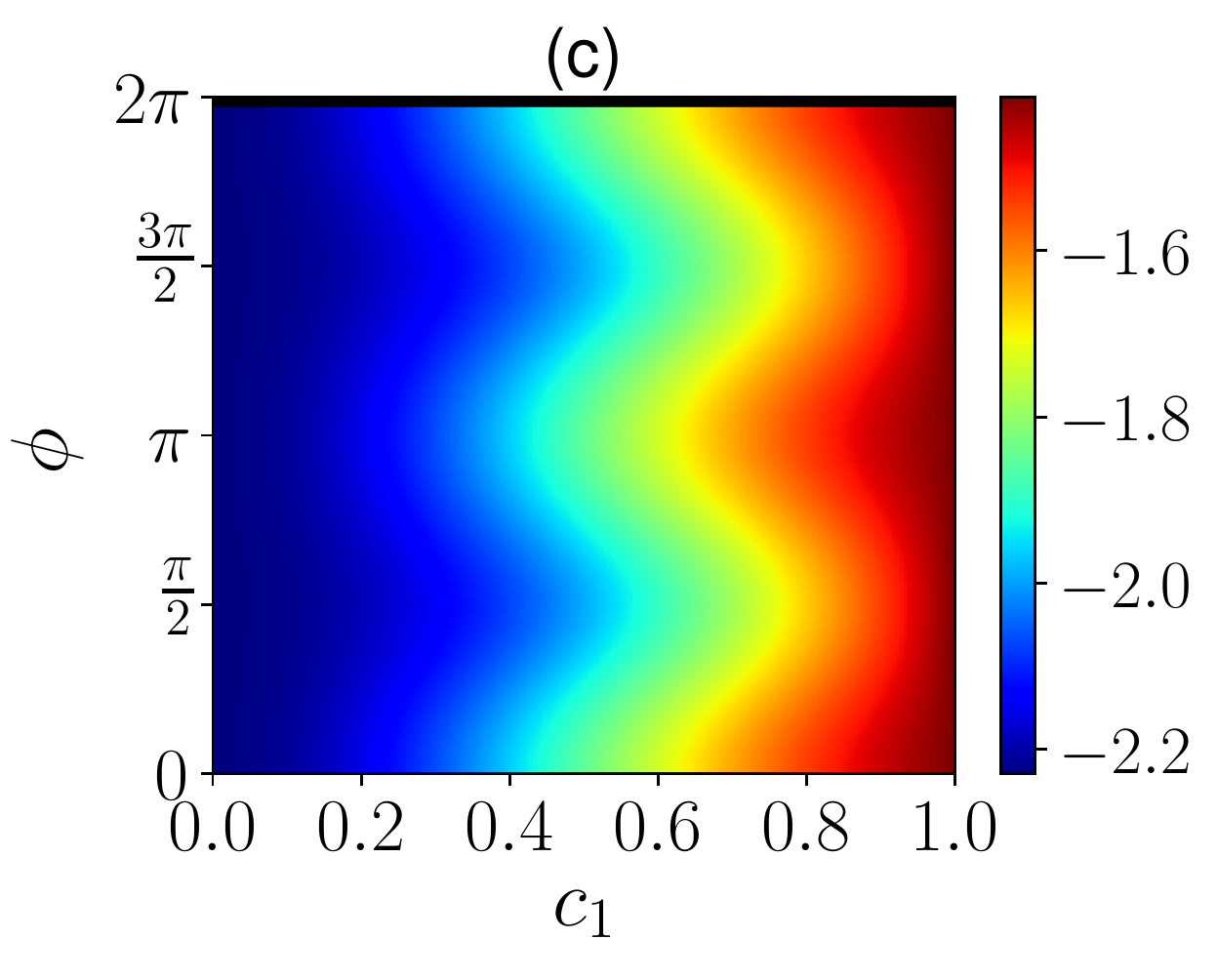}\includegraphics[width=0.25\linewidth]{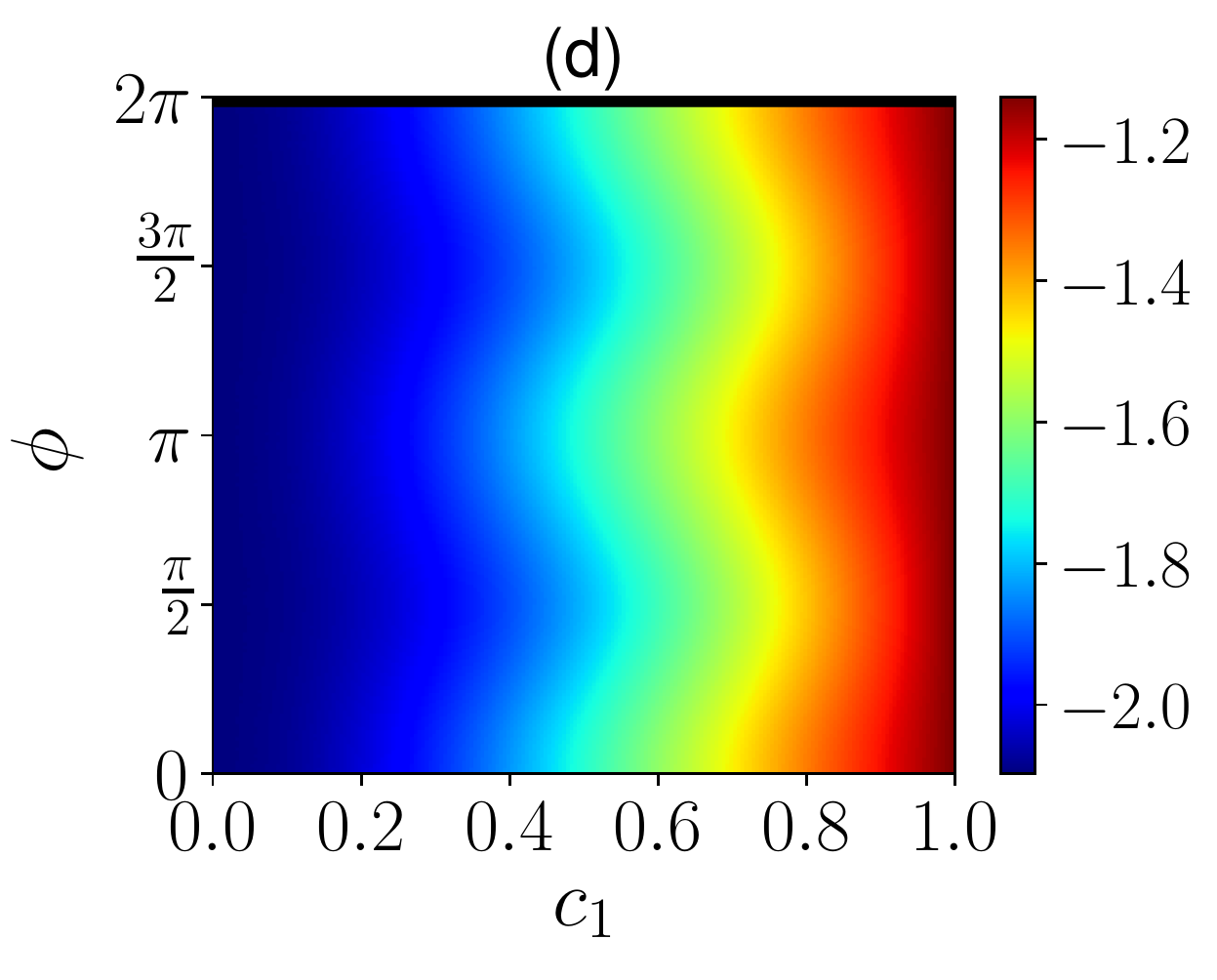} 
\end{center}
\caption{(a) Renyi-2 entropy of the state $\ket{\eta} = c_{1}\ket{\psi_{1}} + c_{2}e^{i\phi}\ket{\psi_{2}}$ depending on $c_{1}$ and $\phi$ at a perturbative strength of $h = 0.15$, 
As the perturbation is increased to $h=0.25$ (b), $h=0.29$ (c), and eventually $h=0.33$ (d), one sees the two minima giving way to a single minimum at $\phi=0, \pi$. ($N$=20 spins)}
\label{fig10}
\end{figure*}
In this section, we analyze the condensing case using Minimally Entangled States (MES). MES are very useful in the construction of the modular $S$-matrix, which is a key
signature for topological order. The general outline of constructing the modular $S$ matrix from MES has been discussed in Refs.~\onlinecite{Ashvin2012,Haldane2013,Pollmann2014}.
The idea of the MES is to compute states which are minimally entangled by observing the Renyi-2 entropy of the region which is trapped inside
a non-trivial cut. In the case of cylinder, we have a single non-trivial cut which encircles the circumference of the cylinder (refer Fig.~\ref{fig13}) and 
we use this to detect the MES, which are two in number for $h=0$ \cite{Ashvin2012}. As the number of Minimally Entangled States in trivial phase is one, we can use the 
change in MES, from two in topologically ordered state to one in the trivial phase, as a signature of phase transition.
For completeness, we sketch the procedure below:
\begin{enumerate}
 \item We start with the linear superposition of ground states, say $\ket{\psi_{1}}$, $\ket{\psi_{2}}$ (starting at $h=0$),
$c_{1}\ket{\psi_{1}} + c_{2}e^{i\phi}\ket{\psi_{2}}$, where $c_{2} = \sqrt{1- c_{1}^{2}}$ and $ 0 \leq c_{1} \leq 1$,  $0 \leq \phi < 2\pi$.
 \item We then find $c_{1}, \phi$ such that the renyi 2-entropy given by $S_{2} = -\log(Tr(\rho_{A}^2))$ is minimized.
 \item We plot the entropy parameterized by $c_{1}, \phi$ and estimate the nature of the entropy.
 \item We observe that the minima occur at $\phi=0, \pi$, so effectively we can minimize the entropy w.r.t $c_{1}$ either for $\phi=0$ or $\phi=\pi$.
 \item We repeat the above step for different system sizes and compute the critical strength in the thermodynamic limit by extrapolating $\frac{1}{R}$ versus $h$ in the limit of $R \rightarrow \infty$.
\end{enumerate}

\begin{figure*}[t]
\begin{center}
\includegraphics[width=0.33\linewidth]{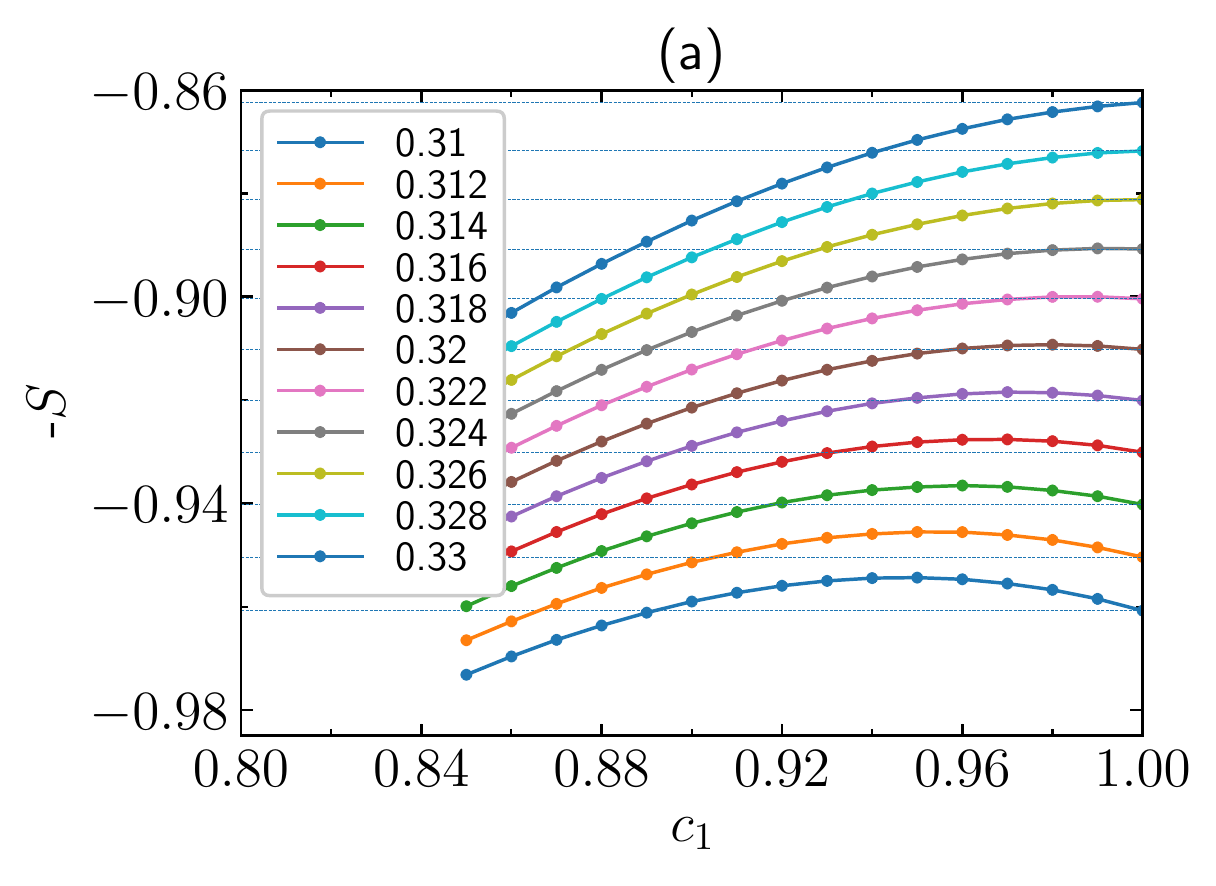}\includegraphics[width=0.33\linewidth]{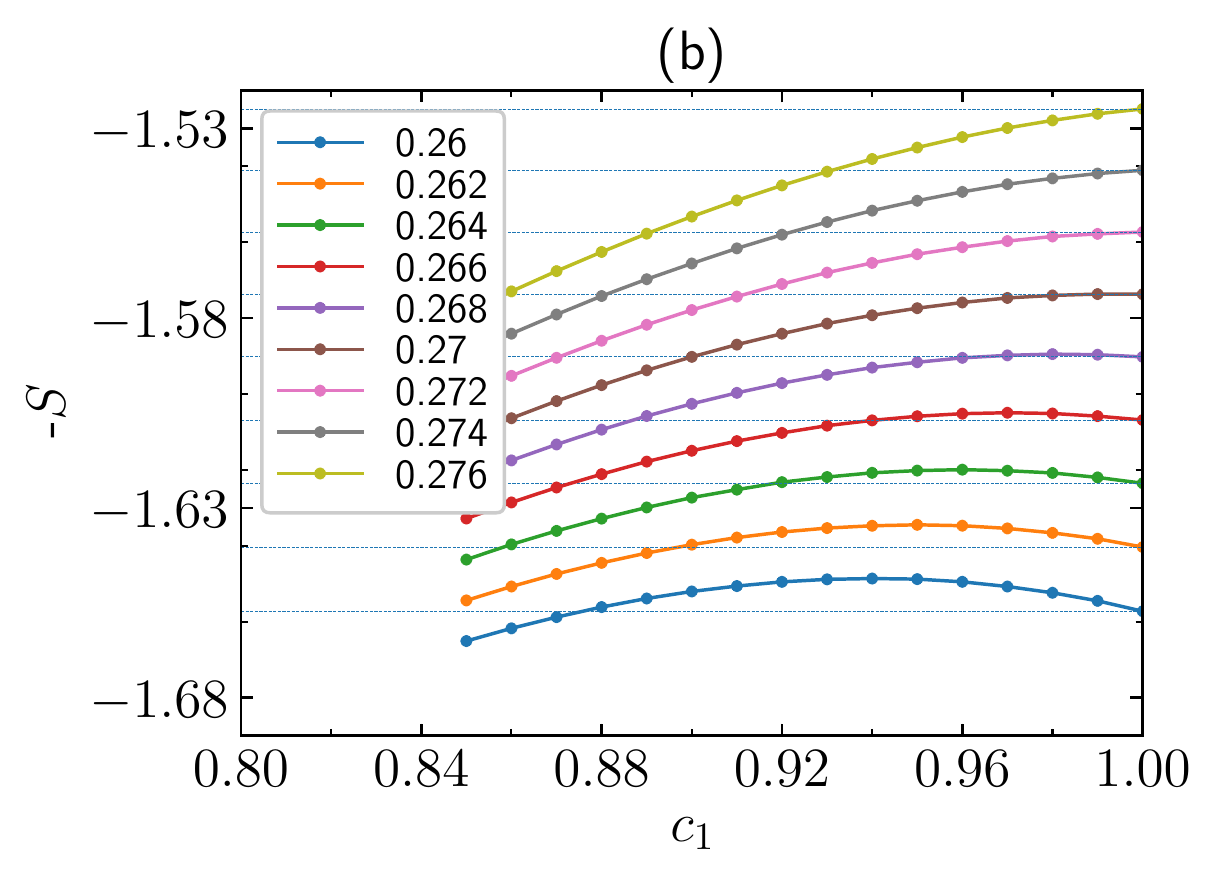}\includegraphics[width=0.33\linewidth]{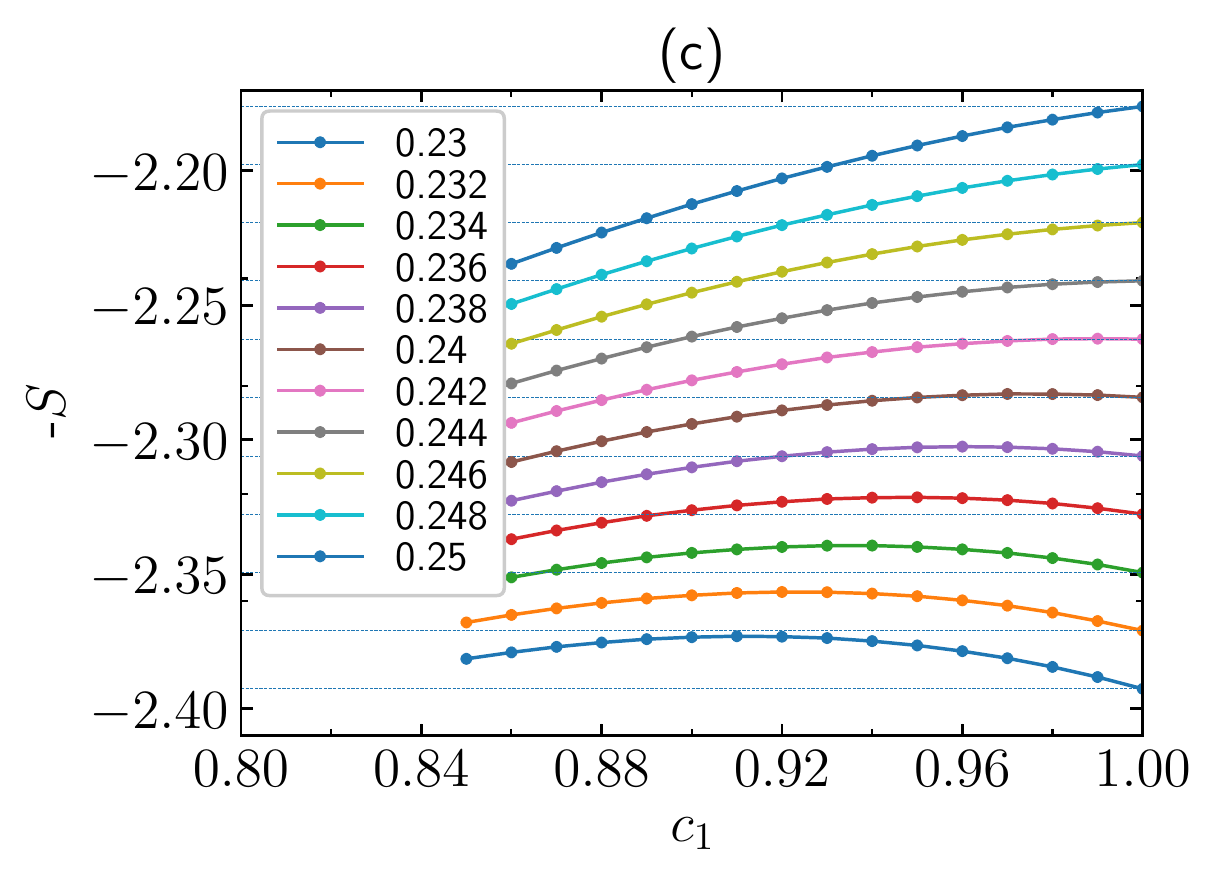}
\end{center} 
\caption{Renyi-2 entropy versus $c_{1}$ at different strengths, captured for different lattice sizes 15 (a), 20 (b), 25 (c). In the case where 
the entropy at $c_{1}$=1 forms a tangent to the entropy versus $c_{1}$ curve (computed at different strength) we denote the critical strength at which the second minimum disappears.}
\label{fig11}
\end{figure*}

We perform the above procedure for the condensing Hamiltonian $H_{idpz}$. As discussed above, we then compute the renyi-2 entropy of the wavefunction, $c_{1}\ket{\psi_{1}} + c_{2}e^{i\phi}\ket{\psi_{2}}$, 
either for $\phi=0$ or $\phi=\pi$ and  minimize w.r.t $c_1$. Clearly, we can see the minimum shifting to $c_1=1$, as in Fig.~\ref{fig10}, which implies that eventually we end up with just one state 
which further implies that the topological order is broken. By observing Fig.~\ref{fig11}, we arrive at a critical strength $h$, for the corresponding system size which scales with $R$.

\begin{figure}[h!]
\begin{center}		
\includegraphics[width=0.85\linewidth]{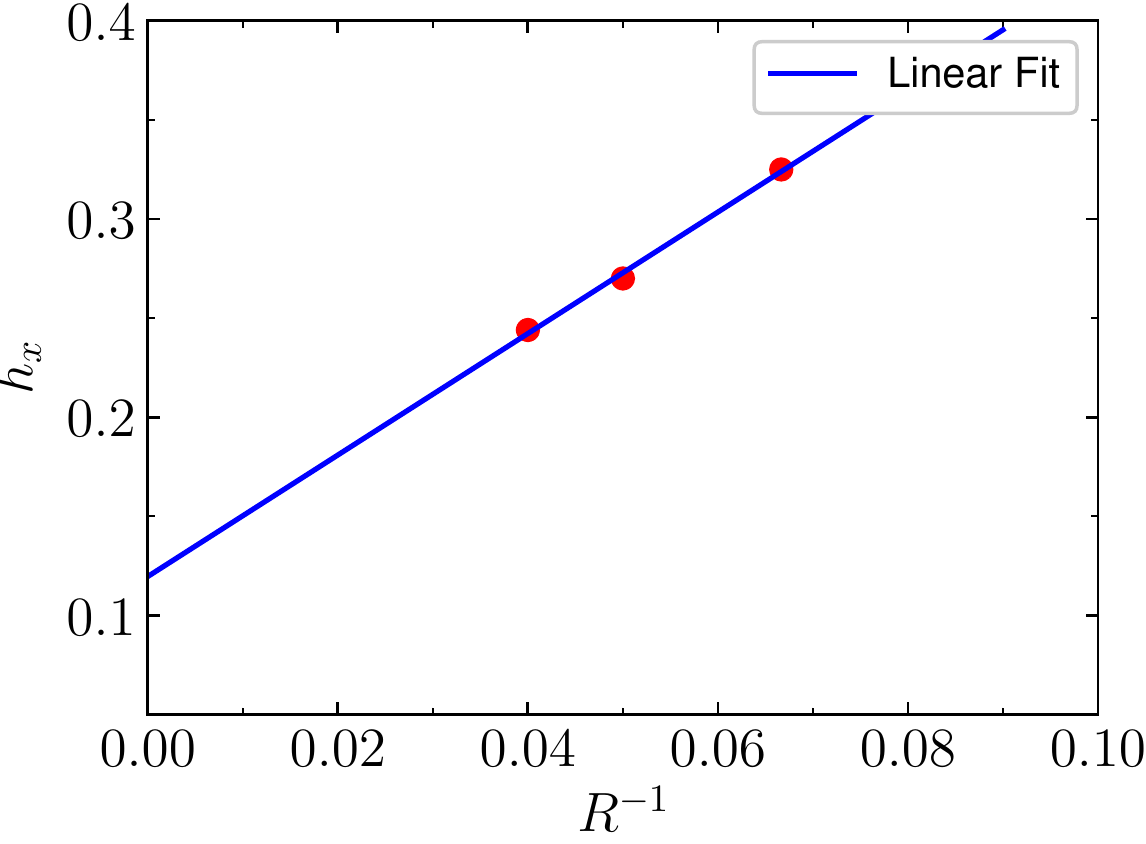}
\end{center} 
\caption{Finite size scaling of the critical strength where the second minimum in the Renyi-2 entropy disappears.}
\label{fig12}
\end{figure}

We infer from Fig.~\ref{fig12}, that critical strength in the thermodynamic limit is given by 0.1196 with an error of $\pm$0.009, which is off from the expected value of zero. We attribute
this error to the computation of the ground states due to the finite size of lattices used, thereby leading to a effective error in the entropy computation. Therefore, we conclude that in this case, 
the MES method fails to properly classify the topological phase transition.

\section{Summary and discussions \label{s6}}
We have analyzed the robustness of topological order for the toric code in an open boundary setting under a perturbation by studying
various properties. We see that the open boundary scenarios with perturbation can be classified into two classes, namely, the condensing and non-condensing classes depending on whether the excitations generated by the perturbation 
get confined or identified at the boundary. Both the condensing and non-condensing classes have been mapped to effective Ising models and thus, these models help provide critical insight into the nature of the phase
transition. To further verify and consolidate the results from the effective Ising models, especially for the condensing case, we have studied the behavior of energy gap, 
TEE and the MES in the exact models. Using the above results we have made an attempt to numerically benchmark the MES method for detecting the phase transition.

In an open boundary setting with a non-condensing scenario, we see that the critical strength occurs at $h=0.453$ and we conclude that topological order in such systems is
more robust when compared to the toric code in a periodic boundary setting under perturbation, which has a critical strength of $h=0.328$ \cite{Nikolay2012}. For the condensing class, the topological 
order breaks as soon as the perturbation is turned on, i.e., for $h>0$ there is no topological order.

In this paper, we have presented robustness under perturbation in an open boundary setting by choosing a cylinder with identical boundary conditions on either 
boundaries. It is also possible to construct a scenario, 
wherein we can identify the open boundaries of the cylinder with different boundary conditions, resulting in a mixed boundary. It would be interesting 
to explore the behavior of other signatures, for example expectation values of Wilson loop operators. It is also possible to interpolate from a toric code on torus to a toric code on cylinder with mixed boundaries by changing the 
underlying topology. We know that both the ground states are topologically ordered and it would be interesting to study different signatures which signal such a phase transition. One other interesting scenario would be to study
the robustness of topological order in systems with domain wall under the effect of perturbation.



\section*{Acknowledgements}
We thank Ling-Yan Hung for fruitful discussions. This
work was funded by the Volkswagen Foundation, by the DFG within SFB
1227 (DQ-mat) and SPP 1929 (GiRyd), by the Thousand Young Talents
 Program, and by the JSPS Grant-In Aid within a JSPS fellowship (P17023).

\appendix

\section{Mapping Toric code in open boundary setting under perturbation to associated Ising models \label{appA}}

\emph{Group as boundary, $K$ = $Z_{2}$, with $\sum_{i}\sigma_{x}^{i}$ as perturbation---}
The Hamiltonian is given by $H_{grpx}$. We note that in the case of group as boundary, the $B_{p}$ violation condenses on the boundary. 
For perturbation, $H_{p} = \sum_{i}\sigma_{x}^{i}$, we note the following relations 
\begin{center}
 $[H_{p}, A_{v}] = 0, \forall v \in bulk$, \\
 $[H_{p}, A_{v^{'}}] = 0, \forall v^{'} \in boundary$, \\
 $[H_{p}, B_{p}] \neq 0$
\end{center}

The perturbation anticommutes with the $B_{p}$ terms and therefore, we shift to the excitation space where
the $B_{p}$ violations are identified at the center of each face by a spin-$\frac{1}{2}$
allowing us to consider the effective Hamiltonian
\be
H_{isgb} = -h_{x}\sum_{p, q}\mu_{p}^{x}\mu_{q}^{x} -h_{x}\sum_{p' \in boundary}\mu_{p'}^{x} - \sum_{p}\mu_{p}^{z}.
\ee
The motivation behind $H_{isgb}$ is that the excitations generated by the perturbation appear in pairs in the bulk and 
are captured by nearest neighbor Ising
interaction, while at the boundary the excitations can exist independently, which is captured by the additional term 
$h_{x}\sum_{p' \in boundary}\mu_{p'}^{x}$. 

The following map captures the essence of the equivalence under the constraint that energy required to 
generate $A_{v}$ violations is set to infinity.
\begin{figure}[h!]
\centering
\begin{tikzpicture}
\node at (-4,0) { $H_{grpx} = \underbrace{\color{black}-\sum_{p}B_{p}} \underbrace{-h_{x}\sum_{i}\sigma_{x}^{i}}$};
\node at (-4,-1.5) {$H_{isgb}$ = $\overbrace{-\sum_{p}\mu_{p}^{z}}\overbrace{-h_{x}\sum_{p, q}\mu_{p}^{x}\mu_{q}^{x} -h_{x}\sum_{p' \in boundary}\mu_{p'}^{x}}$};
\draw[thick, ->] (-4.1,-0.5) -- (-5.75,-0.95);
\draw[thick, ->] (-2.6,-0.5) -- (-2.6,-0.95);
\end{tikzpicture}
\end{figure}

Similar arguments can be constructed for the other Hamiltonian $H_{idpz}$ which results in a condensing scenario. 
Instead of shifting to the excitation space of $B_{p}$, we shift to the excitation space of $A_{v}$ where excitations are identified at vertices with a
spin-$\frac{1}{2}$. 

To summarize, we note that Hamiltonians $H_{idpz}$ and $H_{grpx}$
are equivalent in the Ising picture and correspond to the condensing class, whose Hamiltonian is given by
\be
   H_{ci} = -h_{p}\sum_{i, j}\mu_{i}^{x}\mu_{j}^{x} - h_{p}\sum_{k \in boundary}\mu_{k}^{x} - \sum_{i}\mu_{i}^{z}.
\ee

\emph{Group as boundary, $K$ = $Z_{2}$, with $\sum_{i}\sigma_{z}^{i}$ as perturbation---}
The Hamiltonian is given by $H_{grpz}$. For perturbation $H_{p} = \sum_{i}\sigma_{z}^{i}$, we note the following relations 
\begin{center}
 $[H_{p}, B_{p}] = 0, \forall p$, \\
 $[H_{p}, A_{v}] \neq 0, \forall v \in bulk$, \\
 $[H_{p}, A_{v^{'}}] \neq 0, \forall v^{'} \in boundary$, \\
\end{center}

We observe that $A_{v}$ violations  appear in pairs due to the perturbation
and shifting to the excitation space where $A_{v}$ violations are identified at the vertices by a spin-$\frac{1}{2}$, 
we have the following Ising Hamiltonian
\be
H_{isgb}^{'} = -h_{z}\sum_{v, w}\mu_{v}^{x}\mu_{w}^{x} - \sum_{v}\mu_{v}^{z}.
\ee
 
Again, the motivation for this effective Ising Hamiltonian is that the excitations generated by the perturbation appear in pairs in the bulk and 
are captured by nearest neighbor Ising interaction as above. Here, this nearest neigbhor interaction is preserved at the boundary as the 
excitations are contained at the boundary. Therefore, this type of boundary does not support isolated excitations at the boundary. In the
same way as above, we can capture the equivalence of the models by the following map
\begin{figure}[ht]
\centering
\begin{tikzpicture}
\node at (-4,0) { $H_{grpz} = \underbrace{\color{black}-\sum_{v}A_{v}} \underbrace{-h_{z}\sum_{i}\sigma_{z}^{i}}$};
\node at (-4,-1.5) {$H_{isgb}^{'}$ = $\overbrace{-\sum_{v}\mu_{v}^{z}}\overbrace{-h_{x}\sum_{v, w}\mu_{v}^{x}\mu_{w}^{x}}.$};
\draw[thick, ->] (-4.1,-0.5) -- (-4.3,-0.95);
\draw[thick, ->] (-2.6,-0.5) -- (-2.6,-0.95);
\end{tikzpicture}
\end{figure}

On similar lines, we can analyze the Hamiltonian $H_{idpx}$, where we move to the excitation space of $B_{p}$
and identify each excitation with a spin-$\frac{1}{2}$ positioned at the center of the face.

To summarize, we note that Hamiltonians $H_{idpx}$ and $H_{grpz}$
are equivalent and form the non-condensing class whose Hamiltonian is given by :
\be
H_{nci} = -h_{p}\sum_{i, j}\mu_{i}^{x}\mu_{j}^{x} - \sum_{i}\mu_{i}^{z}.
\ee

\section{CNOT mechanism in the context of open boundaries \label{appB}}
We present the CNOT mechanism which maps the Hamiltonian $H_{grpx}$ to the equivalent Ising picture along with the 
topological spins and vacancy. This section heavily relies on the derivation mentioned in Ref. ~\onlinecite{Vidal2011} and 
we extend the below map on top of the existing literature mentioned in the above reference. 
\newpage
\onecolumngrid

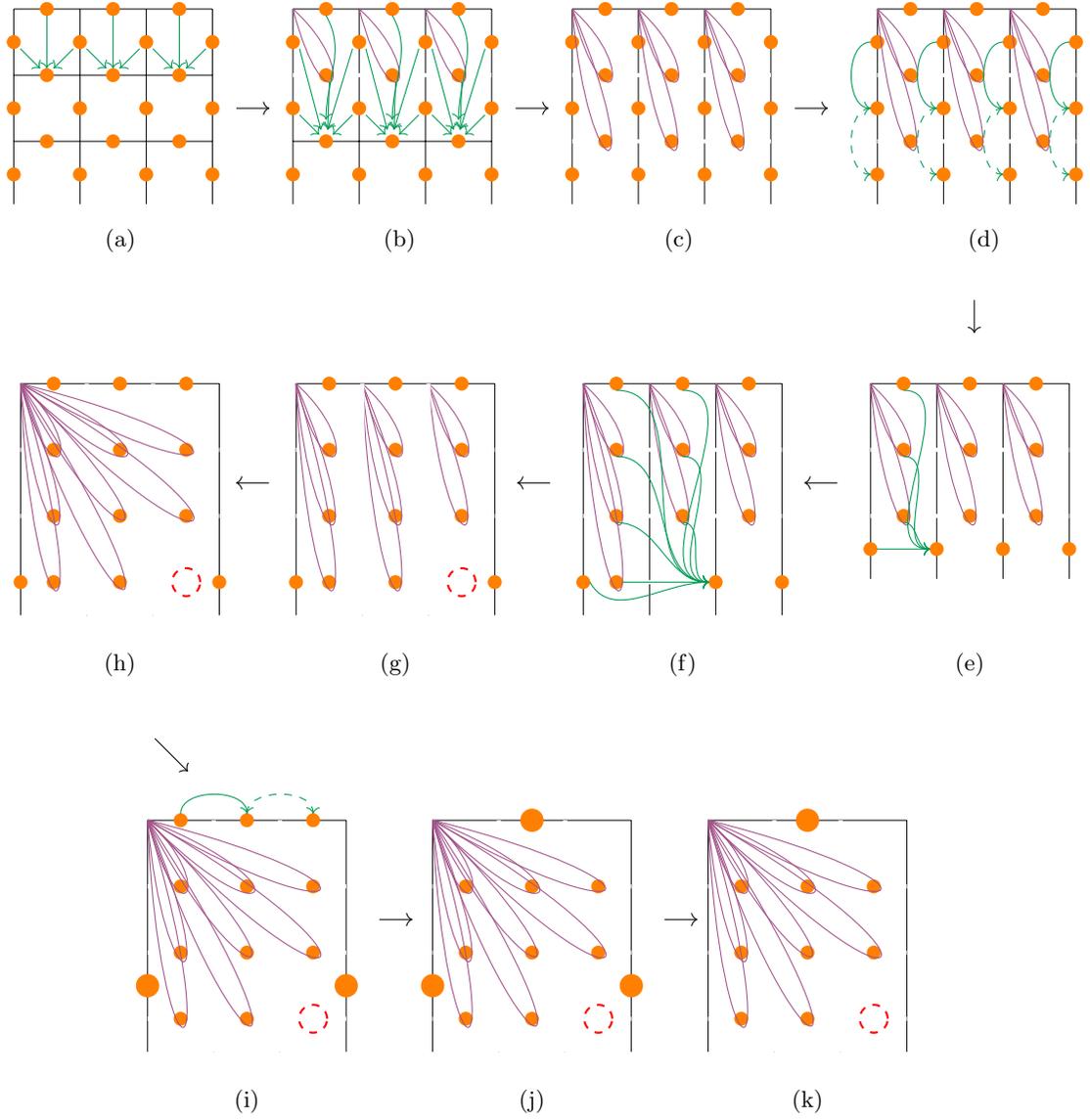
\begin{figure}[h!] 
\begin{minipage}{\linewidth}
\begin{tikzpicture}[scale=0.45]
\draw[ForestGreen, -{>[scale=1.0]}] (0.2,4.8) -- (0.8,4.2); 
\draw[ForestGreen, -{>[scale=1.0]}] (1.,5.8) -- (1.,4.2); 
\draw[ForestGreen, -{>[scale=1.0]}] (1.8,4.8) -- (1.2,4.2);

\draw[ForestGreen, -{>[scale=1.0]}] (2.2,4.8) -- (2.8,4.2); 
\draw[ForestGreen, -{>[scale=1.0]}] (3.,5.8) -- (3.,4.2); 
\draw[ForestGreen, -{>[scale=1.0]}] (3.8,4.8) -- (3.2,4.2);

\draw[ForestGreen, -{>[scale=1.0]}] (4.2,4.8) -- (4.8,4.2); 
\draw[ForestGreen, -{>[scale=1.0]}] (5.,5.8) -- (5.,4.2); 
\draw[ForestGreen, -{>[scale=1.0]}] (5.8,4.8) -- (5.2,4.2);

\draw[step=2.cm,black,very thin] (0.,0.1) grid (6.,6.);
     \foreach \x in {0,2,4,6} 
 		\foreach \y in {1,3,5}
 	  	   \fill[orange] (\x,\y) circle (6pt);
     \foreach \x in {1,3,5} 
 		\foreach \y in {2,4,6}
 	  	   \fill[orange] (\x,\y) circle (6pt);
\hspace{0.1cm}
\draw[-{>[scale=1.0]}] (6.5,3.) -- (7.5,3.);
\node[scale=1.] at (3,-1) {(a)};
\end{tikzpicture}
\hspace{0.1cm}
\begin{tikzpicture}[scale=0.45]
\draw[step=2.cm,black,very thin] (0.,0.1) grid (6.,6.);
\draw[step=2.cm,white,ultra thick] (0.,4.) -- (6.,4.);
     \foreach \x in {0,2,4,6} 
 		\foreach \y in {1,3,5}
 	  	   \fill[orange] (\x,\y) circle (6pt);
     \foreach \x in {1,3,5} 
 		\foreach \y in {2,4,6}
 	  	   \fill[orange] (\x,\y) circle (6pt);

\draw[DarkOrchid] (0,6) .. controls (1.,2.2) and (2.15,4.05) .. (0,6);
\draw[DarkOrchid] (2,6) .. controls (3.,2.2) and (4.15,4.05) .. (2,6);
\draw[DarkOrchid] (4,6) .. controls (5.,2.2) and (6.15,4.05) .. (4,6);

\draw[ForestGreen, -{>[scale=1.0]}] (0.2,2.8) -- (0.8,2.2); 
\draw[ForestGreen, -{>[scale=1.0]}] (0.2,4.8) -- (0.85,2.35); 
\draw[ForestGreen, -{>[scale=1.0]}] (1.8,2.8) -- (1.2,2.2); 
\draw[ForestGreen, -{>[scale=1.0]}] (1.8,4.8) -- (1.05,2.35);
\draw[ForestGreen, -{>[scale=1.0]}] (1,5.8) to[out=-60,in=90] (1,2.65);
\draw[ForestGreen, -{>[scale=1.0]}] (1.1,3.8) -- (1.05,2.35);

\draw[ForestGreen, -{>[scale=1.0]}] (2.2,2.8) -- (2.8,2.2); 
\draw[ForestGreen, -{>[scale=1.0]}] (2.2,4.8) -- (2.85,2.35); 
\draw[ForestGreen, -{>[scale=1.0]}] (3.8,2.8) -- (3.2,2.2); 
\draw[ForestGreen, -{>[scale=1.0]}] (3.8,4.8) -- (3.05,2.35);
\draw[ForestGreen, -{>[scale=1.0]}] (3,5.8) to[out=-60,in=90] (3,2.65);
\draw[ForestGreen, -{>[scale=1.0]}] (3.1,3.8) -- (3.05,2.35);

\draw[ForestGreen, -{>[scale=1.0]}] (4.2,2.8) -- (4.8,2.2); 
\draw[ForestGreen, -{>[scale=1.0]}] (4.2,4.8) -- (4.85,2.35); 
\draw[ForestGreen, -{>[scale=1.0]}] (5.8,2.8) -- (5.2,2.2); 
\draw[ForestGreen, -{>[scale=1.0]}] (5.8,4.8) -- (5.05,2.35);
\draw[ForestGreen, -{>[scale=1.0]}] (5,5.8) to[out=-60,in=90] (5,2.65);
\draw[ForestGreen, -{>[scale=1.0]}] (5.1,3.8) -- (5.05,2.35);

\hspace{0.1cm}
\draw[-{>[scale=1.0]}] (6.5,3.) -- (7.5,3.);
\node[scale=1.] at (3,-1) {(b)};
\end{tikzpicture}
\hspace{0.1cm}
\begin{tikzpicture}[scale=0.45]
\draw[step=2.cm,black,very thin] (0.,0.1) grid (6.,6.);
\draw[step=2.cm,white,ultra thick] (0.,4.) -- (6.,4.);
\draw[step=2.cm,white,ultra thick] (0.,2.) -- (6.,2.);
     \foreach \x in {0,2,4,6} 
 		\foreach \y in {1,3,5}
 	  	   \fill[orange] (\x,\y) circle (6pt);
     \foreach \x in {1,3,5} 
 		\foreach \y in {2,4,6}
 	  	   \fill[orange] (\x,\y) circle (6pt);

\draw[DarkOrchid] (0,6) .. controls (1.,2.2) and (2.15,4.05) .. (0,6);
\draw[DarkOrchid] (2,6) .. controls (3.,2.2) and (4.15,4.05) .. (2,6);
\draw[DarkOrchid] (4,6) .. controls (5.,2.2) and (6.15,4.05) .. (4,6);

\draw[DarkOrchid] (0,6) .. controls (0.5,0.3) and (2.4,.3) .. (0,6);
\draw[DarkOrchid] (2,6) .. controls (2.5,0.3) and (4.4,.3) .. (2,6);
\draw[DarkOrchid] (4,6) .. controls (4.5,0.3) and (6.4,.3) .. (4,6);
\hspace{0.1cm}
\draw[-{>[scale=1.0]}] (6.5,3.) -- (7.5,3.);
\node[scale=1.] at (3,-1) {(c)};
\end{tikzpicture}
\hspace{0.1cm}
\begin{tikzpicture}[scale=0.45]
\draw[step=2.cm,black,very thin] (0.,0.1) grid (6.,6.);
\draw[step=2.cm,white,ultra thick] (0.,4.) -- (6.,4.);
\draw[step=2.cm,white,ultra thick] (0.,2.) -- (6.,2.);
     \foreach \x in {0,2,4,6} 
 		\foreach \y in {1,3,5}
 	  	   \fill[orange] (\x,\y) circle (6pt);
     \foreach \x in {1,3,5} 
 		\foreach \y in {2,4,6}
 	  	   \fill[orange] (\x,\y) circle (6pt);

\draw[DarkOrchid] (0,6) .. controls (1.,2.2) and (2.15,4.05) .. (0,6);
\draw[DarkOrchid] (2,6) .. controls (3.,2.2) and (4.15,4.05) .. (2,6);
\draw[DarkOrchid] (4,6) .. controls (5.,2.2) and (6.15,4.05) .. (4,6);

\draw[DarkOrchid] (0,6) .. controls (0.5,0.3) and (2.4,.3) .. (0,6);
\draw[DarkOrchid] (2,6) .. controls (2.5,0.3) and (4.4,.3) .. (2,6);
\draw[DarkOrchid] (4,6) .. controls (4.5,0.3) and (6.4,.3) .. (4,6);

\draw[ForestGreen, -{>[scale=1.0]}] (-0.2,5) to[out=180,in=180] (-0.2,3);
\draw[ForestGreen, -{>[scale=1.0]}, dashed] (-0.2,3) to[out=180,in=180] (-0.2,1);
\draw[ForestGreen, -{>[scale=1.0]}] (1.8,5) to[out=180,in=180] (1.8,3);
\draw[ForestGreen, -{>[scale=1.0]}, dashed] (1.8,3) to[out=180,in=180] (1.8,1);
\draw[ForestGreen, -{>[scale=1.0]}] (3.8,5) to[out=180,in=180] (3.8,3);
\draw[ForestGreen, -{>[scale=1.0]}, dashed] (3.8,3) to[out=180,in=180] (3.8,1);
\draw[ForestGreen, -{>[scale=1.0]}] (5.8,5) to[out=180,in=180] (5.8,3);
\draw[ForestGreen, -{>[scale=1.0]}, dashed] (5.8,3) to[out=180,in=180] (5.8,1);
\hspace{0.1cm}
\node[scale=1.] at (3,-1) {(d)};
\end{tikzpicture}
\vspace{0.5cm}
\end{minipage}
\begin{minipage}{0.65\linewidth}\flushright
\begin{tikzpicture}[scale=0.45]
\draw[-{>[scale=1.0]}] (4,-3.) -- (4,-4.);
\end{tikzpicture}
\end{minipage}
\begin{minipage}{\linewidth}
\vspace{0.55cm}
\begin{tikzpicture}[scale=0.45]
\draw[step=2.cm,black,very thin] (0.,-1.) grid (6.,6.);
\draw[step=2.cm,white,ultra thick] (0.,4.) -- (6.,4.);
\draw[step=2.cm,white,ultra thick] (0.,2.) -- (6.,2.);
\draw[step=2.cm,white,ultra thick] (0.,0.) -- (6.,0.);
\draw[step=2.cm,white,ultra thick] (2.,-1.) -- (2.,6.);
\draw[step=2.cm,white,ultra thick] (4.,-1.) -- (4.,6.);
     \foreach \x in {0,6} 
 		\foreach \y in {0}
 	  	   \fill[orange] (\x,\y) circle (6pt);
     \foreach \x in {1,3,5} 
 		\foreach \y in {2,4,6}
 	  	   \fill[orange] (\x,\y) circle (6pt);

\draw[DarkOrchid] (0,6) .. controls (1.,2.2) and (2.15,4.05) .. (0,6);
\draw[DarkOrchid] (0,6) .. controls (3.7,2.2) and (4.9,4.05) .. (0,6);
\draw[DarkOrchid] (0,6) .. controls (6.3,2.3) and (7.6,4.1) .. (0,6);

\draw[DarkOrchid] (0,6) .. controls (0.5,0.3) and (2.4,.3) .. (0,6);
\draw[DarkOrchid] (0,6) .. controls (2.75,.5) and (5.45,.5) .. (0,6);
\draw[DarkOrchid] (0,6) .. controls (6.05,-.5) and (7.85,1.5) .. (0,6);

\fill[orange] (1,0) circle (6pt);
\draw[DarkOrchid] (0,6) .. controls (0.35,-2.35) and (2.5,-2.35) .. (0,6);
\fill[orange] (3,0) circle (6pt);
\draw[DarkOrchid] (0,6) .. controls (2.8,-2.3) and (5.45,-2.3) .. (0,6);

\draw[red,thick,dashed] (5,0) circle (12pt);
\node[scale=1.] at (3,-2.5) {(h)};
\draw[-{>[scale=1.0]}] (7.5,3.) -- (6.5,3.);
\end{tikzpicture}
\begin{tikzpicture}[scale=0.45]
\draw[step=2.cm,black,very thin] (0.,-1.) grid (6.,6.);
\draw[step=2.cm,white,ultra thick] (0.,4.) -- (6.,4.);
\draw[step=2.cm,white,ultra thick] (0.,2.) -- (6.,2.);
\draw[step=2.cm,white,ultra thick] (0.,0.) -- (6.,0.);
     \foreach \x in {0,6} 
 		\foreach \y in {0}
 	  	   \fill[orange] (\x,\y) circle (6pt);
     \foreach \x in {1,3,5} 
 		\foreach \y in {2,4,6}
 	  	   \fill[orange] (\x,\y) circle (6pt);

\draw[DarkOrchid] (0,6) .. controls (1.,2.2) and (2.15,4.05) .. (0,6);
\draw[DarkOrchid] (2,6) .. controls (3.,2.2) and (4.15,4.05) .. (2,6);
\draw[DarkOrchid] (4,6) .. controls (5.,2.2) and (6.15,4.05) .. (4,6);

\draw[DarkOrchid] (0,6) .. controls (0.5,0.3) and (2.4,.3) .. (0,6);
\draw[DarkOrchid] (2,6) .. controls (2.5,0.3) and (4.4,.3) .. (2,6);
\draw[DarkOrchid] (4,6) .. controls (4.5,0.3) and (6.4,.3) .. (4,6);

\fill[orange] (1,0) circle (6pt);
\draw[DarkOrchid] (0,6) .. controls (0.35,-2.35) and (2.5,-2.35) .. (0,6);
\fill[orange] (3,0) circle (6pt);
\draw[DarkOrchid] (2,6) .. controls (2.35,-2.35) and (4.5,-2.35) .. (2,6);

\draw[red,thick,dashed] (5,0) circle (12pt);
\draw[step=2.cm,white,ultra thick] (2.,-1.) -- (2.,6.);
\draw[step=2.cm,white,ultra thick] (4.,-1.) -- (4.,6.);
\node[scale=1.] at (3,-2.5) {(g)};
\end{tikzpicture}
\begin{tikzpicture}[scale=0.45]
\draw[step=2.cm,black,very thin] (0.,-1.) grid (6.,6.);
\draw[step=2.cm,white,ultra thick] (0.,4.) -- (6.,4.);
\draw[step=2.cm,white,ultra thick] (0.,2.) -- (6.,2.);
\draw[step=2.cm,white,ultra thick] (0.,0.) -- (6.,0.);
     \foreach \x in {0,4,6} 
 		\foreach \y in {0}
 	  	   \fill[orange] (\x,\y) circle (6pt);
     \foreach \x in {1,3,5} 
 		\foreach \y in {2,4,6}
 	  	   \fill[orange] (\x,\y) circle (6pt);

\draw[DarkOrchid] (0,6) .. controls (1.,2.2) and (2.15,4.05) .. (0,6);
\draw[DarkOrchid] (2,6) .. controls (3.,2.2) and (4.15,4.05) .. (2,6);
\draw[DarkOrchid] (4,6) .. controls (5.,2.2) and (6.15,4.05) .. (4,6);

\draw[DarkOrchid] (0,6) .. controls (0.5,0.3) and (2.4,.3) .. (0,6);
\draw[DarkOrchid] (2,6) .. controls (2.5,0.3) and (4.4,.3) .. (2,6);
\draw[DarkOrchid] (4,6) .. controls (4.5,0.3) and (6.4,.3) .. (4,6);

\fill[orange] (1,0) circle (6pt);
\draw[DarkOrchid] (0,6) .. controls (0.35,-2.35) and (2.5,-2.35) .. (0,6);

\draw[ForestGreen, -{>[scale=1.0]}] (3.,5.8) to[out=0,in=180] (3.8,0);
\draw[ForestGreen, -{>[scale=1.0]}] (3,3.8) to[out=0,in=180] (3.8,0);
\draw[ForestGreen, -{>[scale=1.0]}] (3,1.8) to[out=0,in=180] (3.8,0);

\draw[ForestGreen, -{>[scale=1.0]}] (1.,5.8) to[out=0,in=180] (3.8,0);
\draw[ForestGreen, -{>[scale=1.0]}] (1.,3.8) to[out=0,in=180] (3.8,0);
\draw[ForestGreen, -{>[scale=1.0]}] (1.,1.8) to[out=0,in=180] (3.8,0);
\draw[ForestGreen, -{>[scale=1.0]}] (1.2,0) to[out=0,in=180] (3.8,0);
\draw[ForestGreen, -{>[scale=1.0]}] (0.2,0) to[out=-60,in=180] (3.8,0);
\draw[-{>[scale=1.0]}] (-1,3.) -- (-2.,3.);
\node[scale=1.] at (3,-2.5) {(f)};
\end{tikzpicture}
\begin{tikzpicture}[scale=0.45]
\draw[step=2.cm,black,very thin] (0.,0.1) grid (6.,6.);
\draw[step=2.cm,white,ultra thick] (0.,4.) -- (6.,4.);
\draw[step=2.cm,white,ultra thick] (0.,2.) -- (6.,2.);
     \foreach \x in {0,2,4,6} 
 		\foreach \y in {1}
 	  	   \fill[orange] (\x,\y) circle (6pt);
     \foreach \x in {1,3,5} 
 		\foreach \y in {2,4,6}
 	  	   \fill[orange] (\x,\y) circle (6pt);

\draw[DarkOrchid] (0,6) .. controls (1.,2.2) and (2.15,4.05) .. (0,6);
\draw[DarkOrchid] (2,6) .. controls (3.,2.2) and (4.15,4.05) .. (2,6);
\draw[DarkOrchid] (4,6) .. controls (5.,2.2) and (6.15,4.05) .. (4,6);

\draw[DarkOrchid] (0,6) .. controls (0.5,0.3) and (2.4,.3) .. (0,6);
\draw[DarkOrchid] (2,6) .. controls (2.5,0.3) and (4.4,.3) .. (2,6);
\draw[DarkOrchid] (4,6) .. controls (4.5,0.3) and (6.4,.3) .. (4,6);

\draw[ForestGreen, -{>[scale=1.0]}] (1.,5.8) to[out=0,in=180] (1.8,1);
\draw[ForestGreen, -{>[scale=1.0]}] (1,3.8) to[out=0,in=180] (1.8,1);
\draw[ForestGreen, -{>[scale=1.0]}] (1,1.8) to[out=0,in=180] (1.8,1);
\draw[ForestGreen, -{>[scale=1.0]}] (0.2,1) to[out=0,in=180] (1.8,1);
\draw[-{>[scale=1.0]}] (-1,3.) -- (-2.,3.);
\node[scale=1.] at (3,-2.5) {(e)};
\end{tikzpicture}
\vspace{0.25cm}
\end{minipage}
\begin{minipage}{0.1\linewidth}
\begin{tikzpicture}[remember picture,overlay, scale=0.45]
  \tikzset{xshift=-15cm,yshift=2cm}
\draw[-{>[scale=1.0]}] (3,-3.) -- (4,-4.);
\end{tikzpicture}
\end{minipage}
\begin{minipage}{\linewidth}
\vspace{1.cm}
\begin{tikzpicture}[scale=0.45]
\draw[step=2.cm,black,very thin] (0.,-1.) grid (6.,6.);
\draw[step=2.cm,white,ultra thick] (0.,4.) -- (6.,4.);
\draw[step=2.cm,white,ultra thick] (0.,2.) -- (6.,2.);
\draw[step=2.cm,white,ultra thick] (0.,0.) -- (6.,0.);
\draw[step=2.cm,white,ultra thick] (2.,-1.) -- (2.,6.);
\draw[step=2.cm,white,ultra thick] (4.,-1.) -- (4.,6.);
     \foreach \x in {0,6} 
 		\foreach \y in {1}
 	  	   \fill[orange] (\x,\y) circle (10pt);
     \foreach \x in {1,3,5} 
 		\foreach \y in {2,4,6}
 	  	   \fill[orange] (\x,\y) circle (6pt);

\draw[DarkOrchid] (0,6) .. controls (1.,2.2) and (2.15,4.05) .. (0,6);
\draw[DarkOrchid] (0,6) .. controls (3.7,2.2) and (4.9,4.05) .. (0,6);
\draw[DarkOrchid] (0,6) .. controls (6.3,2.3) and (7.6,4.1) .. (0,6);

\draw[DarkOrchid] (0,6) .. controls (0.5,0.3) and (2.4,.3) .. (0,6);
\draw[DarkOrchid] (0,6) .. controls (2.75,.5) and (5.45,.5) .. (0,6);
\draw[DarkOrchid] (0,6) .. controls (6.05,-.5) and (7.85,1.5) .. (0,6);

\fill[orange] (1,0) circle (6pt);
\draw[DarkOrchid] (0,6) .. controls (0.35,-2.35) and (2.5,-2.35) .. (0,6);
\fill[orange] (3,0) circle (6pt);
\draw[DarkOrchid] (0,6) .. controls (2.8,-2.3) and (5.45,-2.3) .. (0,6);

\draw[red,thick,dashed] (5,0) circle (12pt);
\draw[ForestGreen, -{>[scale=1.0]}] (1.,6.2) to[out=90,in=90] (3.,6.2);
\draw[ForestGreen, -{>[scale=1.0]}, dashed] (3.,6.2) to[out=90,in=90] (5.,6.2);
\draw[-{>[scale=1.0]}] (7.,3.) -- (8.,3.);
\node[scale=1.] at (3,-2.5) {(i)};
\end{tikzpicture}
\begin{tikzpicture}[scale=0.45]
\draw[step=2.cm,black,very thin] (0.,-1.) grid (6.,6.);
\draw[step=2.cm,white,ultra thick] (0.,4.) -- (6.,4.);
\draw[step=2.cm,white,ultra thick] (0.,2.) -- (6.,2.);
\draw[step=2.cm,white,ultra thick] (0.,0.) -- (6.,0.);
\draw[step=2.cm,white,ultra thick] (2.,-1.) -- (2.,6.);
\draw[step=2.cm,white,ultra thick] (4.,-1.) -- (4.,6.);
     \foreach \x in {0,6} 
 		\foreach \y in {1}
 	  	   \fill[orange] (\x,\y) circle (10pt);
     \foreach \x in {1,3,5} 
 		\foreach \y in {2,4}
 	  	   \fill[orange] (\x,\y) circle (6pt);

\fill[orange] (3,6) circle (10pt);

\draw[DarkOrchid] (0,6) .. controls (1.,2.2) and (2.15,4.05) .. (0,6);
\draw[DarkOrchid] (0,6) .. controls (3.7,2.2) and (4.9,4.05) .. (0,6);
\draw[DarkOrchid] (0,6) .. controls (6.3,2.3) and (7.6,4.1) .. (0,6);

\draw[DarkOrchid] (0,6) .. controls (0.5,0.3) and (2.4,.3) .. (0,6);
\draw[DarkOrchid] (0,6) .. controls (2.75,.5) and (5.45,.5) .. (0,6);
\draw[DarkOrchid] (0,6) .. controls (6.05,-.5) and (7.85,1.5) .. (0,6);

\fill[orange] (1,0) circle (6pt);
\draw[DarkOrchid] (0,6) .. controls (0.35,-2.35) and (2.5,-2.35) .. (0,6);
\fill[orange] (3,0) circle (6pt);
\draw[DarkOrchid] (0,6) .. controls (2.8,-2.3) and (5.45,-2.3) .. (0,6);

\draw[red,thick,dashed] (5,0) circle (12pt);
\draw[-{>[scale=1.0]}] (7.,3.) -- (8.,3.);
\node[scale=1.] at (3,-2.5) {(j)};
\end{tikzpicture}
\begin{tikzpicture}[scale=0.45]
\draw[step=2.cm,black,very thin] (0.,-1.) grid (6.,6.);
\draw[step=2.cm,white,ultra thick] (0.,4.) -- (6.,4.);
\draw[step=2.cm,white,ultra thick] (0.,2.) -- (6.,2.);
\draw[step=2.cm,white,ultra thick] (0.,0.) -- (6.,0.);
\draw[step=2.cm,white,ultra thick] (2.,-1.) -- (2.,6.);
\draw[step=2.cm,white,ultra thick] (4.,-1.) -- (4.,6.);
     \foreach \x in {1,3,5} 
 		\foreach \y in {2,4}
 	  	   \fill[orange] (\x,\y) circle (6pt);

\fill[orange] (3,6) circle (10pt);

\draw[DarkOrchid] (0,6) .. controls (1.,2.2) and (2.15,4.05) .. (0,6);
\draw[DarkOrchid] (0,6) .. controls (3.7,2.2) and (4.9,4.05) .. (0,6);
\draw[DarkOrchid] (0,6) .. controls (6.3,2.3) and (7.6,4.1) .. (0,6);

\draw[DarkOrchid] (0,6) .. controls (0.5,0.3) and (2.4,.3) .. (0,6);
\draw[DarkOrchid] (0,6) .. controls (2.75,.5) and (5.45,.5) .. (0,6);
\draw[DarkOrchid] (0,6) .. controls (6.05,-.5) and (7.85,1.5) .. (0,6);

\fill[orange] (1,0) circle (6pt);
\draw[DarkOrchid] (0,6) .. controls (0.35,-2.35) and (2.5,-2.35) .. (0,6);
\fill[orange] (3,0) circle (6pt);
\draw[DarkOrchid] (0,6) .. controls (2.8,-2.3) and (5.45,-2.3) .. (0,6);

\draw[red,thick,dashed] (5,0) circle (12pt);
\node[scale=1.] at (3,-2.5) {(k)};
\end{tikzpicture}
\end{minipage}
\caption{Ising map along with the topological spins (denoted by large orange dots) and the vacancy (denoted by dashed red circle) for 
Hamiltonian $H_{grpx}$. We note that all the above steps are as outlined in Ref. ~\onlinecite{Vidal2011}. The difference between the map in
the periodic case and the open boundary case arises in (f), where the last rung is retained because of the boundary. Step (j) is equivalent to 
(k) since the action of two vertical non-trivial loop operators (homotopic loop operators) on the system have no effective action,  
$L_{topological}^{2} = \mathds{I}$, where $L_{topological}$ is the non-trivial loop operator in the vertical direction. 
Thereby the only loop operator remaining is in the horizontal direction connecting both the boundaries as indicated in (k). The topological 
spin in the horizontal direction couples the Ising spins on either side of the shorter boundary.}
\end{figure}

\twocolumngrid 
\newpage
\mbox{}
\newpage

\newpage
\bibliographystyle{aip}
\bibliography{bib.bib}
\end{document}